\documentclass[11pt]{article}

\usepackage[dvips]{graphicx}
\usepackage{epsfig,amsmath,amssymb,verbatim,mathrsfs,array,layout,textcomp,latexsym, slashed,graphicx,bbm,physics,tensor,cite}

\usepackage[normalem]{ulem}
\usepackage{appendix}
\usepackage{enumitem}
\usepackage{rotating}
\usepackage[hidelinks]{hyperref}
\usepackage[usenames,dvipsnames]{color} 

\usepackage{soul}
\usepackage{dsfont}

\usepackage{tikz}
\usepackage{tikz-feynman}
\tikzset{
	cross/.style={path picture={\draw[black]
			(path picture bounding box.south east) -- (path picture bounding box.north west)
			(path picture bounding box.south west) -- (path picture bounding box.north east);}}
}

\allowdisplaybreaks

\makeatletter

\newenvironment{bottompar}{\par\vspace*{\fill}}{\clearpage}
\newcommand*{\xdash}[1][3em]{\rule[0.5ex]{#1}{0.55pt}}
\def\mysection#1{{\bf #1.} }

\newcommand{\refeq}[1]{(\ref{#1})}
\def\mysection#1{{\bf #1.}}

\newcommand\mydots{\hbox to 1.1em{$\,\cdot\hss\cdot\hss\cdot\,$}}

\begin{document}

\begin{titlepage}
	\setcounter{page}{1} \baselineskip=15.5pt 
	\thispagestyle{empty}
	$\quad$
	{\raggedleft IFT UAM-CSIC 24-168\par}
	\vskip 60 pt
	
	\begin{center}
		{\fontsize{18}{18} \bf Finite parts of inflationary loops}
	\end{center}

	\vskip 20pt
	\begin{center}
		\noindent
		{\fontsize{13}{30}\selectfont  Guillermo Ballesteros$^{1,2}$, Jes\'us Gamb\'in Egea$^{2}$, and Flavio Riccardi$^{1,2}$}
	\end{center}

	\begin{center}
		\vskip 4pt
		\textit{ $^1${Departamento de F\'{\i}sica Te\'{o}rica, Universidad Aut\'{o}noma de Madrid (UAM), \\Campus de Cantoblanco, 28049 Madrid, Spain}
		}
		\vskip 5pt
		\textit{ $^2${Instituto de F\'{\i}sica Te\'{o}rica UAM-CSIC,  Campus de Cantoblanco, 28049 Madrid, Spain}
		}
	\end{center}
	
	\vspace{0.4cm}
	\centerline{\bf Abstract}
	\vspace{0.3cm}
	\noindent 
	
	We present a method for solving loop integrals in dimensional regularization that is particularly useful in the context of inflation. We apply this method to the calculation of the tensor power spectrum induced by scalar fluctuations in slow-roll inflation.

	\begin{bottompar}
		\noindent\xdash[15em]\\
		\small{
			guillermo.ballesteros@uam.es\\
			j.gambin@csic.es\\
			flavio.riccardi@uam.es}
	\end{bottompar}
\end{titlepage}

\setcounter{page}{2}
\newpage

\tableofcontents

\section{Introduction} \label{sec:Intro}

The study of Quantum Field Theory (QFT)  observables in inflationary cosmologies has gained increasing attention over the past two decades, particularly following Maldacena's seminal calculation of the three-point correlations of scalar and tensor perturbations in slow-roll inflation at leading order (tree-level) in interactions \cite{Maldacena:2002vr}. 
The standard tool for computing such objects in inflation is the in-in formalism, initially introduced by Schwinger \cite{Schwinger:1960qe}, Bakshi \& Mahanthappa \cite{Bakshi:1962dv,Bakshi:1963bn} and Keldysh \cite{Keldysh:1964ud}, and later applied in a cosmological context by Calzetta and Hu \cite{Calzetta:1986ey}.
Weinberg provided a useful reformulation of the in-in formalism \cite{Weinberg:2005vy,Weinberg:2006ac}, offering both a path integral and an operator approach, which has been shown to be a powerful tool for computing correlation functions of inflationary primordial fluctuations, see e.g.\ \cite{Woodard:2014jba,Carlip:2015asa,Chen:2010xka}. The main strength of the in-in operator formalism lies in the validity of perturbation theory, allowing predictions to be made through a systematic expansion in small fluctuations.

The smallness of large scale primordial fluctuations in standard inflationary scenarios may lead to deem the calculation of cosmological correlators beyond the lowest perturbative order unimportant. However, as it was noted in \cite{Weinberg:2005vy,Weinberg:2006ac}, their study can provide insights 
into the properties of inflation as a QFT, even if observational verification of these higher-order contributions may be elusive with observations of the Cosmic Microwave Background (CMB) and other probes of the early universe. 	Moreover, calculations beyond the naive leading order are relevant in certain contexts.
For instance, in cases where the tree-level contribution to an observable quantity is suppressed, or even vanishes due to symmetry arguments, higher-order corrections (loops) are required. 
This may be the case of gravitational waves induced during inflation by scalar modes in concrete scenarios. 
Additionally, the study of loop corrections may help to verify the validity of perturbation theory. For instance, it was argued in \cite{Kristiano:2022maq} that the one-loop scalar power spectrum at CMB scales in models featuring an ultra slow-roll inflation phase could signal its breakdown. This claim has been challenged, see e.g.\ \cite{Riotto:2023hoz,Franciolini:2023lgy,Tada:2023rgp,Firouzjahi:2023bkt,Inomata:2024lud,Ballesteros:2024zdp,Fumagalli:2024jzz}, with \cite{Ballesteros:2024zdp} arguing that shorter scales should be used instead.

The logarithmic ultraviolet (UV) divergences and the logarithmic finite contributions to the one-loop scalar power spectrum of curvature fluctuations were obtained in \cite{Senatore:2009cf}, correcting the earlier result of \cite{Weinberg:2005vy}. Two regularization methods (dimensional regularization and a cutoff) were used in \cite{Senatore:2009cf}, finding consistency between both. In the context of inflation, it is important to take into account that the interactions and the dynamics of the free fields are modified by the change in the number of dimensions when dimensional regularization is applied.\footnote{The need of changing the free modes and the interactions according to the number of dimensions was already noted in \cite{Weinberg:2005vy}.} In the case of cutoff regularization, a physical one (as opposed to comoving) should be used to deal with momentum and time integrals. 

A regulator that breaks a symmetry gives results that in general do not respect that symmetry. It is therefore often convenient to use regularization methods that respect the symmetries of a problem. One such method is dimensional regularization. 
The loop integrals appearing with dimensional regularization in the context of inflation are considerably more complicated than for scattering processes in flat spacetime.
So far, only in relatively simple cases these integrals have been solved (see e.g.\ \cite{Chen:2016nrs}) and in general only logarithmic contributions are extracted. This motivates the necessity of finding a systematic procedure that allows to obtain these integrals in dimensional regularization. Doing so, we go one step further in the understanding of the renormalization of the in-in observables in cosmology.  We present a method to obtain the complete divergent and finite parts of the loop integrals in dimensional regularization. 
Although our method is generally applicable to any type of integral in dimensional regularization, we apply it in a specific example: the tensor power spectrum, $\mathcal{P}_h$, induced by scalar fluctuations at one-loop in inflation.

In the next section we present the method at the core of this paper. In Section \ref{sec:Action} we discuss the action of the example we analyze. Specifically, Section~\ref{sec:Interactions} is devoted to the free dynamics of tensor and scalar fluctuations during inflation at lowest order in slow-roll, as well as the set of scalar-tensor interactions relevant for the calculation of $\mathcal{P}_h$. And Section \ref{sec:Action cts} discusses the counterterms needed for renormalization. Then, in Section \ref{sec:Tensor spectrum} we obtain the functional form of the tree-level, one-loop and counterterm contributions to $\mathcal{P}_h$, with Section \ref{sec:Tensor Ph in SR inf} presenting the regularized one-loop $\mathcal{P}_h$. 
In Section \ref{sec:Cutoff} we reconsider $\mathcal{P}_h$, this time regularizing with a cutoff, obtaining the same result as with dimensional regularization after restoring the symmetries broken by the cutoff.
Finally, we summarize our results in Section \ref{sec:Conclusions}. The paper also contains some appendices. 
In Appendix \ref{ap: Bubbles and disconnected parts} we discuss bubble diagrams, showing that their contribution to general $n$-point correlation functions vanishes in the in-in formalism, as well as that $n$-point correlation functions are constructed from connected correlation functions of lower order. 
In Appendix~\ref{sec:QED} we compare our results for $\mathcal{P}_h$ to the well-known case of vacuum polarization in 
	quantum electrodynamics 
(QED). 
In Appendix \ref{ap:in-out} we discuss the relation between the renormalization of $\mathcal{P}_h$ in the in-in formalism and the in-out formalism.

\section{Dimensional regularization procedure for loop integrals}
\label{sec:dimregprocedure}

In QFT, UV divergences typically arise in the computation of observable quantities at loop level.  
However, the UV physics can be modified by means of regularization, in such a way that the result for observables is finite \cite{Schwinger:1948iu, Schwinger:1949zz, Feynman:1948fi, Tomonaga:1948zz, Dyson:1949bp}. Although this may seem to introduce an unavoidable dependence on the choice of regularization method, the counterterms in the action used to absorb the divergences can remove this dependence. 
A common UV regulator is a cutoff. 
However, this regulator breaks the symmetries of the theory.
For instance, as we will later see in the case of the tensor power spectrum induced by scalar fluctuations during inflation, using a cutoff as a regulator forces us to introduce a mass counterterm for the tensor modes. 
A method of regularization that respects the symmetries is dimensional regularization \cite{Bollini:1972bi,tHooft:1972tcz,Cicuta:1972jf,Ashmore:1972uj} (see also e.g.\ \cite{Senatore:2009cf,Chen:2016nrs} for a discussion in the inflationary context).\footnote{In cases where the symmetries are related to the number of dimensions, dimensional regularization may not respect these symmetries and some modifications are required (see e.g.\ \cite{Siegel:1979wq,Siegel:1980qs,Capper:1979ns}).} Dimensional regularization isolates 
the relevant divergences, i.e.\ the logarithmic ones (see e.g.\ \cite{tHooft:1972tcz}), which are the ones that leave a running after renormalization.\footnote{Logarithmic divergences are usually accompanied by (logarithmic) finite parts that cannot be absorbed by counterterms.}$^,$\footnote{A priori, regularization does not necessarily imply that renormalization is possible and, in principle, one should check that the adequate counterterms exist.}

  The idea behind it consists in changing the dimensionality of spacetime in such a way that the convergence of the loop integrals is improved. This can be intuitively understood considering the simplest integration measure in $d$ dimensions: $\dd^d p \sim p^{d-1} \dd p$. By gradually making $d$ small enough,  convergence is improved in the UV, eventually reaching UV finiteness.

Providing analytical expressions for the integrals commonly found in dimensional regularization is, in general, not easy. Even more so in a dynamical spacetime, where the time evolution of the free fields can be complicated. In this section we describe a procedure that allows to obtain these integrals, extracting both the finite and divergent parts. 

Let us consider a dimensional-regularized one-loop calculation\footnote{A generalization of the procedure to higher-order loops can be implemented.} 
characterized as the following integral:
\begin{equation} \label{eq:I delta definition}
	I(\delta) = \int _0^\infty\dd p \: p^\delta f(\delta,p)\,.
\end{equation}
In order to regularize this integral with dimensional regularization, we must assume that there is some natural number $N$ such that $|f(\delta,p)|\le C \, p^N$ in the limit $p\to\infty$, for some positive constant $C$.
The number of 
 dimensions considered is $d = (3+\delta)$, so that the term $p^\delta$ comes from the
volume element 
of the loop integral. The function $f(\delta,p)$ {--which we have defined in such a way that it contains the factor $p^2$ from the measure--} comes from the time and angular integrals that arise in the in-in formalism. Thus, $f(\delta,p)$ is determined by the particular observable being considered and on the free modes as well as the structure of the interactions. Of course, both $f(\delta,p)$ and $I(\delta)$ are functions of the external momenta, the masses of the particles and in general of any energy scale and couplings intervening. We do not indicate explicitly these dependencies for the sake of brevity in the notation. By construction, no terms like $p^\delta$ (i.e.~$ 1 + \delta \log p $, making a leading order expansion in $\delta$ around $\delta = 0$) can arise within the function $f(\delta,p)$ due to the type of objects that compose it. Thus, we reiterate that the only $p^\delta$ term is the one coming from the volume element. In general, $f(\delta,p)$ will have a dependence on $p$ such that, prior to regularization, it would give rise to a UV divergent contribution to $I(\delta)$. As we discussed above, this is where, thanks to the term $p^\delta$ and choosing $\delta$ small enough (even negative), the convergence of the integral improves to the point of the latter becoming finite.

In what follows, we assume that $I(\delta)$ has no infrared (IR) divergences for simplicity.
However, the procedure we present for dealing with UV divergences via dimensional regularization can be generalized to the case with IR divergences.
Since the only source of divergence is in the UV, we split the integral as follows:
\begin{equation} \label{split}
	I(\delta) = \int _0^L\dd p \: p^\delta f(\delta,p) + \int _L^\infty\dd p \: p^\delta f(\delta,p)\,,
\end{equation}
where $L$ is an arbitrary, finite comoving scale. The first integral is already finite without any need of regularization,
so there we can take $\delta =0$, i.e.\
\begin{equation}\label{eq:Int dim. reg.}
	I(\delta) = \int _0^L \dd p \: f(0,p) + \int _L^\infty\dd p \: p^\delta f(\delta,p) +\mathcal{O}(\delta) \,,
\end{equation}
where the notation $\mathcal{O}(\delta)$ simply indicates that we are neglecting terms in a Taylor expansion around $\delta = 0$ that are at least linear in $\delta$ (and hence vanish when  $\delta \rightarrow 0$).  
The regulator $\delta$ introduces a modification of the UV part of the integral $I(0)$	
whose purpose is to make the result finite.

The strategy to solve the second integral is
to take the limit $p\to \infty$ of {the integrand $p^\delta f(\delta,p)$,} keeping only the terms that give rise to divergent or oscillatory contributions to $I(\delta)$. The convergent terms will give corrections that vanish in the limit where $L$ is arbitrarily large, which will be the one we will take for convenience. We will assume that $\delta$ acquires the necessary value so that the integral converges at the upper limit of the momentum integral, as explained above. Once the UV divergences of the second integral are gone, we analytically continue $\delta$ in the complex plane. Finally, we can return to the case of interest by taking the limit $\delta\to 0$.

Although this procedure can be reminiscent 
of the method of regions \cite{Beneke:1997zp} (see also \cite{Smirnov:1999bza,Jantzen:2011nz}), there the expansions of the integrand $p^\delta f(\delta,p)$ are made in different regimes of $p$, maintaining in each of these expansions the complete integration regime (i.e.\ the whole range of $p$ is integrated) as well as the dependence on the regulator (i.e.\ $\delta$ is not taken to be zero).  
The method of regions is useful for simplifying certain loop integrals in flat spacetime, but
can be
complicated to apply it in 
the context of inflation	
(see \cite{Beneke:2023wmt} for an example of one such application). Our method, however, is nothing more than a literal application of dimensional regularization, but as far as we know it has not been presented in this form before.

\subsection{Analysis of particular cases}

Let us assume $f(\delta,p)$ to be a holomorphic function in a neighborhood of $p=\infty$ with the exception of a possible isolated pole at $\infty$. Then it  can be expanded in a Laurent series around $\infty$ in the second integral in eq.\ (\ref{split}). Moreover, we also assume that infinity is not an essential singularity, i.e.\ the Laurent series has a finite number of terms at infinity: 
\begin{equation} \label{eq:UV behavior easy}
	f(\delta, p) = \sum_{n=-\infty}^N p^n c_n(\delta)\,,
\end{equation}
where $c_n(\delta)$ are dimensionful coefficients which may depend on the external momenta, and which are analytic in $\delta = 0$.
The second integral of eq.\ \eqref{eq:Int dim. reg.} is
\begin{equation}
	\int _L^\infty\dd p \: p^\delta f(\delta,p) = \int _L^\infty\dd p \: p^\delta \sum_{n=-\infty}^N p^n c_n(\delta)= \sum_{n=-\infty}^N \frac{p^{\delta + n + 1}}{\delta + n + 1}\eval_L^\infty \: c_n(\delta)\,.
\end{equation}
To ensure the UV convergence of this integral, we have to assume that $\Re \{\delta + N + 1 \}<0$. Thus:
\begin{equation}
	\int _L^\infty\dd p \: p^\delta f(\delta,p) = - \sum_{n=-\infty}^N \frac{L^{\delta + n + 1}}{\delta + n + 1}\: c_n(\delta).
\end{equation}
Once the integral has been regularized, obtaining a result that depends on $\delta$, we can complexify the result by taking it as valid for all values of $\delta\in\mathbb{C}$, including those values of $\delta$ that would not allow the integral to be previously convergent, e.g.\ $\delta = 0$. We can then take the limit $\delta\to0$, taking special care with the term $n = -1$,
\begin{equation}
	\int _L^\infty\dd p \: p^\delta f(\delta,p) = -\frac{c_{-1}(0)}{\delta} - \frac{\dd c_{-1}}{\dd \delta}\bigg|_{\delta=0} - c_{-1}(0)\log L- \sum_{\substack{n=-\infty \\ n\ne -1}}^N \frac{L^{n + 1}}{n + 1}\: c_n(0) + \order{\delta}\,.
\end{equation}
Then, eq.\ \eqref{eq:Int dim. reg.} becomes:
\begin{equation}
	I(\delta) = \int _0^L \dd p \: f(0,p) - c_{-1}(0)\log L- \sum_{\substack{n=-\infty \\ n\ne -1}}^N \frac{L^{n + 1}}{n + 1}\: c_n(0)-\frac{c_{-1}(0)}{\delta} - \frac{\dd c_{-1}}{\dd \delta}\bigg|_{\delta=0} +\order{\delta}\,.
\end{equation}
This equation holds $\forall L >0$ and, as expected, the dependence on $L$ of the first integral is canceled with the remaining terms coming from the second integral. To prove it, we note that $\partial_L I(\delta) = 0$, where we have used eq.\ \eqref{eq:UV behavior easy}. This arbitrariness in the choice of $L$ allows us to send $L\to \infty$, making all the terms with $n<-1$ in the sum over $n$ vanish:
\begin{equation}
	\label{eq:Finite parts dim reg}
		\begin{aligned}
			I(\delta) = \lim_{L\to \infty} \left( \int _0^L \dd p \: f(0,p) - c_{-1}(0)\log L- \sum_{n=0}^N \frac{L^{n + 1}}{n + 1}\: c_n(0)\right) -\frac{c_{-1}(0)}{\delta} - \frac{\dd c_{-1}}{\dd \delta}\bigg|_{\delta=0} +\order{\delta}\,.
	\end{aligned}
\end{equation}
This result makes clear that dimensional regularization isolates the logarithmic divergences, grouped in the coefficient $c_{-1}(0)$. We also note that the $\delta$-dependence of $f(\delta,p)$ manifests in the appearance of {${\dd c_{-1}}/{\dd \delta}|_{\delta=0}$.} 
Thus, the integral $I(\delta)$ is only sensitive to the information of $f(\delta,p)$ and its first derivative $\partial_\delta f(\delta,p)$ in $\delta = 0$. We will explore the implications of this result later.

Although eq.\ \eqref{eq:UV behavior easy} is one of the most representative cases, in practice we may encounter cases where the UV limit of $f(\delta,p)$ is very different. 
It is worth considering scenarios in which the UV behavior of $f(\delta,p)$ is oscillatory, having an essential singularity at infinity. Such cases will be encountered repeatedly in the calculation of observables during inflation, where the UV limit of $f(\delta,p)$ involves phases. In general, the effect of the phase improves the convergence of the integrals, so that an a priori logarithmic divergence, $f(\delta,p) \sim e^{i p / p_*} / p$ for $p\to \infty$, does not introduce a single pole in $\delta = 0$. 
However, terms like $f(\delta,p) \sim e^{i p / p_*} $ for $p\to \infty$ give a UV finite but oscillatory --in the momentum $p$-- contribution to the integrals, and are therefore a priori problematic. As we are going to see next, the general procedure described above for the application of dimensional regularization correctly deals with this type of behavior. 

Let us consider a function $f$ that for $p\to \infty$ has the asymptotic behavior of a phase times a Laurent series with no essential singularity at infinity, i.e:
\begin{equation}
f(\delta,p) = \, e^{i p / p_*}\sum_{n=-\infty}^N c_n(\delta)p^n,
\end{equation} 
then:
\begin{equation}
	I(\delta) = \int_0^L \dd p\, f(0,p) + \sum_{n=-\infty}^N c_n(\delta) \int_L^\infty \dd p \, p^{\delta + n} e^{i p / p_*} + \order{\delta}\,.
\end{equation}
For all the integrals in the sum to be convergent in the UV, we must assume that $\Re{\delta + N}<0$,
\begin{equation}
	I(\delta) = \int_0^L \dd p\, f(0,p) + \sum_{n=-\infty}^N c_n(\delta) \, L^{1+\delta + n} {\rm E}_{-n-\delta}(-i L/p_*)+ \order{\delta}\,, \quad {\rm where} \quad {\rm E}_n(z) = \int_1^\infty \dd x \frac{e^{-z\, x}}{x^n}\,.
\end{equation}  
We can take the limit $\delta \to 0$ without finding poles in $\delta = 0$,\footnote{The absence of single poles in $\delta = 0$, contrary to what is obtained in eq.\ \eqref{eq:Finite parts dim reg}, is a manifestation of the fact that the limits $p_*\to \infty$ and $\delta \to 0$ are non-commutative.} obtaining:
\begin{equation}
\label{eq:finiteL}
	I(\delta) = \int_0^L \dd p\, f(0,p) +\sum_{n=-\infty}^N c_n(0) \, L^{1 + n} {\rm E}_{-n}(-i L/p_*) + \order{\delta}\,.
\end{equation} 
Given the asymptotic behavior for $L\to\infty$ of $\left|\mathrm{E}_{-n}\left(-iL/p_*\right)\right|\to|p_*| /L$, and since eq.\ \eqref{eq:finiteL} holds $\forall L>0$, we can send $L\to\infty$ canceling all the negative terms in the sum:
\begin{align}
I(\delta) = \lim_{L\to\infty}\left(\int_0^L \dd p\, f(0,p) +\sum_{n=0}^N c_n(0) \, L^{1 + n} {\rm E}_{-n}(-i L/p_*)\right) + \order{\delta}\,.
\end{align}
Using the representation of the function ${\rm E}_{-n}$ valid for $n\in \mathbb{N}$:
\begin{align}
{\rm E}_{-n}(z) = n!\, e^{-z}\sum_{k=0}^n \frac{z^{k-n-1}}{k!},
\end{align}
and exchanging the sums using the property $\sum_{n=0}^N \sum_{k=0}^n A_{nk}=\sum_{k=0}^N \sum_{n=k}^N A_{nk}$, we find:
\begin{equation}
\label{eq:Finite parts dim reg with phases}
I(\delta) = \lim_{L\to\infty}\left(\int_0^L \dd p\, f(0,p) +e^{iL/p_*}\sum_{k=0}^N \frac{L^k}{k!(ip_*)^k} \sum_{n=k}^N c_n(0) n!\left(ip_*\right)^{n+1}\right) + \order{\delta}\,.
\end{equation}

We must be especially careful when logarithms $\sim \log p$ are involved --violating the hypothesis of holomorphicity at infinity-- since, as opposed to phases, they worsen the convergence of the integrals.
	However, since the dimensional regularization procedure we are applying is general, we can continue to use it in this case too. Indeed, let us consider, for example, the term $1/p \subset f(\delta,p)$ which a priori gives rise to a simple pole in $\delta = 0$, but we will accompany it with some power, $n \in \mathbb{N}$, of a logarithm, i.e.\
\begin{equation}
	\int_L^\infty \dd p \: \frac{p^\delta}{p} \log^n p =  \frac{(-1)^{n+1} n!}{\delta^{n+1}} - \frac{\log^{n+1} L}{n+1} + \order{\delta} \,,
\end{equation}
being manifest that the degree of divergence in $\delta$ increases when $n$ does. Although it is important to characterize such cases, in practice {it seems unlikely to  encounter them in the context of inflation. The function} $f(\delta,p)$ will depend on the free fields (satisfying Bunch-Davies initial conditions) and the structure of the interactions, which in general will not give rise to this type of behavior. However, there are situations where such higher-order poles in $\delta$ can arise, e.g.\ in the presence of non-local interactions involving operators such as $\log \partial^2$. We will not discuss such situations here.

\subsection{Simplified method: expansion around $3$ spatial dimensions} \label{sec:dimreg simplified}

The procedure described above is often not practical due to the complexity of $f(\delta,p)$ as a function of $\delta$. 
However, as we have seen above, assuming that $f(\delta,p)$ is holomorphic at $p\to \infty$, $I(\delta)$ will only involve single poles in $\delta = 0$.\footnote{To be precise, we also need to assume that $f(\delta, p)$ is holomorphic in $\delta=0$, which in practice is no more than assuming that the coefficients $c(\delta)$ arising in the expansion of $f(\delta,p)$ in $p\to \infty$ do not involve poles in $\delta = 0$. This is motivated by the fact that in the limit $\delta \to 0$ we recover the original theory in $3$ spatial dimensions.} Therefore, we can do the expansion:
\begin{equation}
	f(\delta, p) = f(0, p) + \delta \, f_{\delta}(0, p) + \order{\delta^2}\quad {\rm where} \quad f_\delta(\delta,p) = \frac{\partial}{\partial \delta}f(\delta,p)\,.
\end{equation}
The contribution to $I(\delta)$ of the $\order{\delta^2}$ terms of this expansion vanishes when the limit $\delta \to 0$ is taken. Therefore, we can identify $c_{-1}(0)$ and ${\dd c_{-1}}/{\dd \delta}\eval_{\delta = 0}$ in eq.\ \eqref{eq:Finite parts dim reg} with the coefficients $\order{p^{-1}}$ of $f(0,p)$ and $f_{\delta}(0,p)$ in the $p\to\infty$ expansion, respectively. 

It is necessary to be careful with the expansion around $\delta = 0$ of certain problematic objects.
In particular, we might ask why the {factor} that guarantees the convergence of $I(\delta)$ in dimensional regularization, $p^\delta$, cannot be expanded. The reason has been answered above: an expansion of $p^\delta$ involves $\log p$. 
Although, as it has been explained, this is not a problem since the described procedure is able to deal with this {kind of structure,} we will not be able to truncate the expansion of $p^\delta$ to linear order since higher-order terms introduce higher-order poles. We will illustrate this problem with a practical example:
\begin{equation} \label{eq:log div expanded}
	\int_L^\infty \dd p \: p^\delta \frac{1}{p} = \int_L^\infty \dd p \: \sum_{n = 0}^\infty \frac{\log^n p}{n!\:p} \delta^n = \sum_{n = 0}^\infty \frac{\log^{n+1} p}{(n+1)!} \delta^n \eval_L^\infty = \frac{p^\delta -1 }{\delta} \eval_L^\infty = -\frac{L^\delta}{\delta}\,,
\end{equation}
assuming $\delta<0$, recovering the expected result, but losing the usefulness of the expansion around $\delta = 0$ since it has been necessary to include all the terms.
We therefore conclude that we can always make the expansion of the integrand by avoiding expanding the terms involving objects that increase the degree of divergence in $\delta$, e.g.\ $p^\delta$ giving rise to $\log p$. As we have commented,  $f(\delta,p)$ {is not expected to involve such}
logarithms.

\subsection{Constants and Fourier volume element in dimensional regularization} \label{sec:Constants in dim reg}

Let us consider the integral:
\begin{equation}
	I(\delta) = \int _0^\infty\dd p \: p^\delta C(\delta) f(\delta,p) =  \int _0^\infty\dd p \: p^\delta C(0) f(\delta,p) + \delta \, \frac{\partial C}{\partial \delta}\eval_{\delta = 0} \int _0^\infty\dd p \: p^\delta f(\delta,p)+\order{\delta}\,,
\end{equation}
where $C(\delta)$ is a dimensionless constant that does not depend on any variable, e.g.\ $(2\pi)^{-3-\delta}$. 
Assuming that $f(\delta,p)$ is holomorphic at $p\to \infty$, due to the factor $\delta$ multiplying the second integral, the finite contribution of the latter to $I(\delta)$ comes from the logarithmic divergence in eq.\ \eqref{eq:Finite parts dim reg},
\begin{equation}
	I(\delta)  =  \int _0^\infty\dd p \: p^\delta C(0) f(\delta,p) -c_{-1}(0) \: \frac{\partial C}{\partial \delta}\eval_{\delta = 0} + \order{\delta}\,.
\end{equation}
The divergent part of $I(\delta)$, together with the contribution of the second integral, will be
\begin{equation}
	I(\delta) \supset -c_{-1}(0) \left( \frac{1}{\delta} + \frac{\partial C}{\partial \delta}\eval_{\delta = 0}\right) \,.
\end{equation}
Assuming that the counterterms can absorb the divergence $1/\delta$, they will also be able to absorb the contribution ${\partial C}/{\partial \delta}\eval_{\delta = 0}$. Therefore, we can choose {a} renormalization scheme such that the counterterms absorb all the effects generated by the dependence of the constants on $\delta$, similar to what is done in the $\overline{\text{MS}}$ scheme in flat space QFT. Henceforth, all the constants will be taken in $\delta = 0$ for simplicity.\footnote{Although this argument applies to global constants of observable quantities, in practice we will take $\delta\to0$ in all the constants of the system. Special care must be taken if UV divergences arise in different diagrams which cannot be absorbed by counterterms but which cancel out once all the contributions are taken into account.}

We will now analyze the Fourier volume element in $d=(3+\delta)$ spatial dimensions, 
\begin{align}
	d^d p = p^{d-1} (\sin \theta_1)^{d-2} (\sin \theta_2)^{d-3}\mydots \sin\theta_{d-2} \, \dd p \, \dd\theta_1\, \dd\theta_2\mydots \dd\theta_{d-2}\, \dd \theta_{d-1},
\end{align}
where we can identify $\theta_1$ and $\theta_2$ with the polar ($\theta$) and azimuthal ($\phi$) angles in 3 dimensions, respectively.
In cases where the integrand only depends on $p$ and $\theta$ we can integrate over the remaining angles and take $\delta = 0$, since the result will be a constant that depends on $\delta$ and as we have seen this effect can be absorbed by the finite part of the counterterms. In these cases, we will replace the volume element with:
\begin{equation}
	\dd^{3 + \delta} p \ \to \  p^{2+\delta} \left( \sin\theta\right) ^{1+\delta} \dd p\,  \dd \theta \, \dd \phi\,.
\end{equation}

\section{Action for scalar and tensor fluctuations} \label{sec:Action}

We are going to illustrate the regularization procedure presented in the previous section with a simple example in the context of inflation: tensor fluctuations induced by scalar ones at zeroth order in the slow-roll expansion.\footnote{Slow-roll corrections, which are not necessary for our purposes, can be added in a straightforward way making some of the ensuing expressions lengthier.}

It is well known that scalar fluctuations during inflation induce second order tensor modes \cite{Matarrese:1992rp, Matarrese:1993zf, Matarrese:1997ay, Nakamura:2006rk, Baumann:2007zm}. This is simply due to the inherently nonlinear nature of gravity. In standard slow-roll inflation, these induced modes are suppressed with respect to the first order tensor ones. However, in models of inflation in which the scalar fluctuations are large enough, the second order effects can become dominant (without breaking perturbation theory). This happens, for instance, {for gravitational waves induced during radiation domination} in models of inflation featuring a period of ultra slow-roll \cite{Kinney:2005vjs} in which the curvature fluctuations at specific scales can be much larger than those inferred from CMB observations, see e.g. \cite{Saito:2009jt}. It is also possible to think of scenarios in which inflationary loops lead to significant tensor or scalar power spectra, which emphasizes the need to accurately calculate the finite contribution from UV divergent loops, as we are going to explore next.

\subsection{Interactions}\label{sec:Interactions}

Let us consider single field inflation and, as we will explain below, work at zeroth order in slow-roll for the resulting spectrum of tensor fluctuations. In practice, this means assuming a de Sitter background evolution. To study the set of interactions that determine the coupling of tensor and scalar fluctuations, we restrict our analysis for simplicity
to the case of an inflaton, $\phi(x)$, minimally coupled to gravity,
\begin{equation}
	S = \frac{1}{2}\int \sqrt{-g} \:\dd^4x \bigg\{ M_P^2 R - g^{\mu\nu}\partial_\mu \phi \partial_\nu \phi - 2V(\phi)\bigg\}\,,
\end{equation}
where the signature of the metric is $(-,+,+,+)$ and $M_P = 1/\sqrt{8\pi G}$ denotes the reduced Planck mass. We use natural units, i.e.\ $\hbar = c = 1$. Greek indices indicate spacetime ones, while Latin indices denote spatial ones.

We use the ADM formalism \cite{Arnowitt:1962hi} to describe the metric
\begin{align}
	ds^2 = -N^2 \dd t^2 + \gamma_{ij} \left(N^i\dd t + \dd x^i \right) \left(N^j\dd t + \dd x^j \right)\,,
	\label{eq:ADM_Decomposition}
\end{align}
where the {\it lapse $N$} and the {\it shift $N^i$} are non-dynamical quantities acting as Lagrange multipliers in the action, and will therefore be determined by the physical degrees of freedom. Following \cite{Maldacena:2002vr}, we use the following scalar-vector-tensor decomposition of the metric:
\begin{align}
	\gamma_{ij} = a^2 \left[ e^\Gamma\right]_{ij}\,,\quad  \Gamma_{ij} = 2\zeta \delta_{ij} + \partial_{ij} E + \partial_{(i} E_{j)} + h_{ij}\,,\quad
	N = 1 + \alpha\,,\quad N_i = \partial_i \beta + \beta_i\,,
	\label{eq:SVT_Decomposition}
\end{align}
where $a$ is the scale factor of the Universe, $\partial^iE_i =\partial^i\beta_i = 0$ and $\partial^ih_{ij} = \tensor{h}{^i_i} = 0$. We also decompose the inflaton into a homogeneous background $\phi_0$ and fluctuations:
\begin{align} \label{decphi}
	\phi(x) = \phi_0(t) + \delta\phi(t,\vb{x})\,.
\end{align}
We use this decomposition of the metric because it is the one that guarantees that, in the $\zeta$-gauge (i.e.\ making $\delta\phi = E = E_i = 0$), both scalar and tensor modes are frozen outside the horizon and at the end of inflation in single field inflation at a classical level \cite{Maldacena:2002vr,Weinberg:2003sw,Lyth:2004gb}. See also \cite{Pimentel:2012tw,Assassi:2012et,Senatore:2012ya,Green:2024fsz} for generalizations of this statement at the quantum level. 

In this gauge and at linear order, $\zeta$ coincides with the gauge invariant quantity $\zeta - H\delta\phi / \dot{\phi}_0$ \cite{Bardeen:1980kt}.
We choose to work instead in the $\delta\phi$-gauge, imposing that $\zeta = E = E_i = 0$, because then the set of non suppressed interactions between the scalar and tensor sector stems solely from the kinetic term of the inflaton \cite{Ballesteros:2024zdp}. The other interactions arising from the metric, through the algebraic variables $N$ and $N^i$, are suppressed by powers of the slow-roll parameter $\epsilon$,\footnote{The $\epsilon$-suppressed interactions are relevant only at the very end of inflation, where by definition $\epsilon = 1$. We neglect them because we are interested in the strict limit $\epsilon\ll1$ for the sake of simplicity.} that follows the usual definition $\epsilon=-\dot{H}/H^2$ in terms of the Hubble rate $H=\dot{a}/a$, where the dot means derivative with respect to cosmic time $t$. Moreover, the tensor fluctuations obtained in the $\delta\phi$-gauge coincide with those of $\zeta$-gauge at the end of inflation for modes with wavenumber $k\ll aH$, since they are not modified under a transformation between these gauges, see e.g.\ \cite{Maldacena:2002vr,Ballesteros:2024zdp}. The action at leading order in $\epsilon$ describing the free dynamics of scalar and tensor fluctuations, as well as the interactions involved in the induction of tensor modes by scalars at second order, is:\footnote{Since we are interested in analyzing only the tensor fluctuations induced by scalars at one loop, we do not consider tensor self-interactions.}$^,$\footnote{There is a quartic interaction term coming from the metric that is not suppressed by powers of $\epsilon$ \cite{Ballesteros:2024zdp},
\begin{equation}
	S\supset \int \dd \tau \dd^3\vb{x}\, \frac{3a^2}{8} \partial^{-2} \partial_i \left( h'_{jk} \partial_i h_{jk} \right) \partial^{-2} \partial_l \left( \delta\phi' \partial_l \delta\phi \right)\,.
\end{equation}
However, its effect on the one-loop tensor spectrum vanishes because the scalar sector comes under a total spatial derivative and $\partial_l \left\langle \delta\phi' \partial_l \delta\phi \right\rangle =0$.} 
\begin{align} \label{eq:Action for the fluctuations}
	\nonumber
	S = \int \dd \tau \dd^3 \vb{x}\, a^2(\tau) \Bigg\{& \frac{M_P^2}{8} \left( (h'_{ij})^2 - (\partial_k h_{ij})^2 \right) +\frac{1}{2} \left(\left( \delta\phi'\right) ^2- (\partial_i\delta\phi)^2 - a^2(\tau) \frac{\partial^2 V(\phi)}{\partial \phi^2}\eval_{\phi_0} \delta\phi^2 \right)\\
	&+ \frac{1}{2} h_{ij} \partial_i\delta\phi \partial_j\delta\phi -\frac{1}{4} h_{ik} h_{kj} \partial_i \delta\phi \partial_j \delta\phi \Bigg\}
	\, ,
\end{align}
where here and through the rest of the text we use squares to denote spatial contractions whenever possible (e.g.\ $(h_{ij})^2 \equiv h_{ij}h_{ij}$)  and  $'$ stands for the derivative with respect to conformal time, $' = \dd / \dd \tau$, being $\dd t = a \dd \tau$. For our purposes, the term containing the second derivative of the potential is negligible; see \cite{Ballesteros:2024zdp} for its expression in terms of slow-roll parameters. However, it is important to stress that the action described in \eqref{eq:Action for the fluctuations} only assumes that $\epsilon \ll 1$, and so it is not restricted to slow-roll inflation, see also \cite{Ballesteros:2024zdp}.

\subsection{Counterterms} \label{sec:Action cts}

The interactions between the tensor and scalar sectors, described by eq.\ \eqref{eq:Action for the fluctuations}, generate a one-loop contribution on the tensor two-point correlation,
\begin{equation}
	\expval{h_{ij}(x)h_{ij}(y)} = 
	\begin{tikzpicture}[baseline={-2}]
		\draw[decorate, decoration=snake] (0,0) -- (2,0);
		\fill[black] (0,0) circle (1.5pt);
		\fill[black] (2,0) circle (1.5pt);
		\node[black] at (0,-0.3) {$x$};
		\node[black] at (2,-0.3) {$y$};
		
		\node[black] at (2.5,0) {$+$};
		
		\draw[decorate, decoration={snake, amplitude=-2pt, segment length=10pt}] (3,0) -- (5,0);
		\draw (4,0.58) circle (0.5);
		\fill[black] (3,0) circle (1.5pt);
		\fill[black] (5,0) circle (1.5pt);
		
		\node[black] at (5.5,0) {$+$};
		
		\draw[decorate, decoration=snake] (6,0) -- (7,0);
		\draw[decorate, decoration=snake] (8,0) -- (9,0);
		\draw (7.5,0) circle (0.5);
		\fill[black] (6,0) circle (1.5pt);
		\fill[black] (9,0) circle (1.5pt);
		
		\node[black] at (6.5,0.3) {$h$};
		\node[black] at (7.5,0.7) {$\delta\phi$};
	\end{tikzpicture}\,,
\end{equation}
that is UV divergent, as usual. To obtain a finite result, which can be related to an observable quantity, a process of regularization and renormalization is required
\cite{Schwinger:1948iu, Schwinger:1949zz, Feynman:1948fi, Tomonaga:1948zz, Dyson:1949bp}; see \cite{Weinberg:2005vy, Seery:2007we, Dimastrogiovanni:2008af, Senatore:2009cf, Weinberg:2010wq,Chen:2016nrs} for previous studies of renormalization in the context of inflation. To proceed, we need to determine the set of counterterms involved in the renormalization of the two-point tensor correlation. Let us analyze the Planck mass suppression of the possible one-loop contributions. Given that the tensor modes scale as $h_{ij}\propto M_P^{-1}$ (see eq.\ \eqref{eq:Action for the fluctuations}), the quartic vertex is proportional to $M_P^{-2}$, whereas the cubic is proportional to $M_P^{-1}$. Therefore, the two types of diagrams that appear at one loop scale in the same way:
\begin{equation}
	\begin{tikzpicture}[baseline={-2}]\draw[decorate, decoration={snake, amplitude=-2pt, segment length=10pt}] (3,0) -- (5,0);
		\draw (4,0.58) circle (0.5);
		\fill[black] (3,0) circle (1.5pt);
		\fill[black] (5,0) circle (1.5pt);
		
		\node[black] at (5.5,0) {$+$};
		
		\draw[decorate, decoration=snake] (6,0) -- (7,0);
		\draw[decorate, decoration=snake] (8,0) -- (9,0);
		\draw (7.5,0) circle (0.5);
		\fill[black] (6,0) circle (1.5pt);
		\fill[black] (9,0) circle (1.5pt);	
	\end{tikzpicture}
	\sim \frac{1}{M_P^4}\,,
\end{equation}
and the counterterms we are looking for must contain the same order in $M_P$, so we require that
\begin{equation}
	\begin{tikzpicture}[baseline={-2}]
		\draw[decorate, decoration=snake] (0,0) -- (2,0);
		\draw[fill=white,cross] (1,0) circle (0.2);
		\fill[black] (0,0) circle (1.5pt);
		\fill[black] (2,0) circle (1.5pt);
	\end{tikzpicture}
	\sim \frac{1}{M_P^4}\,.
\end{equation}
{Cutting the external legs ($\propto M_P^{-2}$), we can see that the coupling constant of the counterterms cannot contain any $M_P$ power. The counterterm Lagrangian necessarily goes as $(h_{ij})^2$ ($\propto M_P^{-2}$).} 

To obtain these counterterms, we use the general covariant action, before expanding in fluctuations \cite{Donoghue:1994dn,Ruhdorfer:2019qmk}. In this way we look for terms such as $R^2$ or $\phi^2R$ that are diffeomorphism invariant and can lead to the counterterms we need. 
First, we note that the counterterms do not involve scalar fluctuations, and therefore as far as the search for counterterms is concerned, we can take $\delta\phi =0$.
The only elements of the metric from which we can obtain a tensor mode are $\gamma_{ij}$, $N$ and $N_i$. Since the scalar, vector and tensor degrees of freedom do not mix at linear order, the variables defined in eq.\ \eqref{eq:SVT_Decomposition} into which the lapse and shift are decomposed do not contain tensor modes, i.e.\ $\alpha = \beta = \beta_i = 0$ at first order in fluctuations (recall that we are taking $\delta\phi =0$). At second order these variables may contain tensor modes, but they are  spatial boundary terms (see e.g.\ the Appendix B of \cite{Ballesteros:2024zdp}) that do not contribute to the tensor two-point correlation. Therefore, we can take $N = 1$ and $N_i = 0$, so that tensor modes may come only from $\gamma_{ij}$. 

Neither the scalar kinetic term, nor the scalar potential, nor $\sqrt{-g}$ depend on $h_{ij}$ (taking $\delta\phi = 0$). Therefore, in order to obtain the counterterms we need to study the Riemann tensor and possible non-minimal couplings of $\phi$ with the metric.
However, since $\phi = \phi_0 \sim \sqrt{\epsilon}$, any non-minimally coupled terms --such as $\phi^2 R$-- will be $\epsilon$ suppressed with respect to those coming purely from the Riemann tensor.\footnote{Inverse powers of $\phi$ cannot give rise to the adequate $M_P$ scaling, which is the only relevant scale for this power counting.}

Taking the considerations above into account, the (diffeomorphism invariant) action that can give rise to the counterterms we are looking for is:
\begin{equation} \label{eq:Cts action general}
	S = \int \sqrt{-g} \,\dd^4x\, \big\{  A_1 R^2 + A_2 R_{\mu\nu}R^{\mu\nu} + A_3 R_{\mu\nu\rho\sigma}R^{\mu\nu\rho\sigma}  + A_4 \square R \big\}\,,
\end{equation}
where we have included terms with up to two extra derivatives (with respect to the Einstein-Hilbert action). The $A_i$ are dimensionless constants. Combinations of these constants will be responsible for absorbing the divergences of the one-loop contribution to the two-point function.  Higher-order terms in curvature, suppressed by powers of $M_P$, do not enter into the one-loop computation of the tensor two-point correlation, but they would be needed at higher loops or to renormalize higher-order correlators. To obtain the explicit form of the counterterms, we expand eq.~\eqref{eq:Cts action general} in fluctuations and keep only quadratic terms:
\begin{align}
	\sqrt{-g}\,R^2 & \supset 6\,a^2 H^2 \left((h'_{ij})^2 - (\partial_k h_{ij})^2 \right) \,,\\
	\sqrt{-g}\,R_{\mu\nu}R^{\mu\nu} & \supset \frac{3}{2}a^2 H^2 \left((h'_{ij})^2 - (\partial_k h_{ij})^2 \right) \,,\\ 
	\sqrt{-g}\,R_{\mu\nu\rho\sigma}R^{\mu\nu\rho\sigma} & \supset 3 a^2 H^2 (h'_{ij})^2 - 4\,a H h'_{ij} \partial^2 h_{ij}  - a^2 H^2 (\partial_k h_{ij})^2 	  - 2\, (\partial_k h'_{ij})^2 +	2\, (\partial^2 h_{ij})^2\,,\\
	\sqrt{-g}\,\square R & \supset \partial_\mu \left(\sqrt{-g}\,\partial^\mu R \right) = \frac{1}{4} \frac{\dd}{\dd \tau}\left[a^2\frac{\dd}{\dd \tau} \left(\frac{1}{a^2} (h'_{ij})^2  - \frac{1}{a^2} (\partial_k h_{ij})^2 \right)\right]  \,,
\end{align}
where we have used $\epsilon \ll 1$ and the free equation of motion of the tensor modes,\footnote{We can use the free equation of motion because, using the in-in formalism to compute correlation functions, the fields evolve in the interaction picture: they follow the dynamics dictated by the free action. Although the equations of motion in $3+\delta$ {spatial} dimensions differ from those in $3$ dimensions, and therefore in dimensional regularization these counterterms will be different, the differences will be removed by the finite part of the counterterms.}
\begin{equation}
	h''_{ij} + 2 a H  h'_{ij} - \partial^2 h_{ij} = 0.
\end{equation}
Since there is no spatial boundary {in de Sitter,}
{terms that are total spatial derivatives}
do not contribute to the correlation functions. This can be shown using momentum conservation at each interaction vertex.  However, this is not the case for temporal boundary terms, whose role can be decisive in this type of calculation \cite{Arroja:2011yj,Burrage:2011hd,Braglia:2024zsl}. Using that
\begin{equation} \label{eq:Not important BT}
	\int \dd \tau \dd^3\vb{x}\, a H h'_{ij} \partial^2 h^{ij} =  \frac{1}{2} \int \dd \tau \dd^3\vb{x}\,\left\{ \frac{\dd}{\dd \tau} \left(- a H (\partial_k h_{ij})^2 \right) + a^2 H^2 (\partial_k h_{ij})^2 \right\}\,,
\end{equation}
where this particular temporal boundary term does not contribute to the two-point correlation function because it does not involve the velocity of $h_{ij}$ (see e.g.\ \cite{Arroja:2011yj}), we finally arrive at the linearly independent action for the counterterms:
\begin{align} \label{eq:Cts general cov}
	S_{\rm cts} = - \int \dd \tau \dd^3\vb{x}\,\bigg\{ C_1 a^2H^2 (h'_{ij})^2  + C_2 a^2 H^2 (\partial_k h_{ij})^2  +C_3\left( (\partial_k h'_{ij})^2 - (\partial^2 h_{ij})^2 \right)  \bigg\}\,,
\end{align}
{where $C_i$ are dimensionless constants (that can be expressed as combinations of the constants $A_i$).}
If we had chosen to take into account only the counterterms coming from $R^2$ and $R_{\mu\nu}R^{\mu\nu}$, we would have obtained the constraint $C_2 = -C_1$. It is the counterterm $\square R$ that breaks this constraint. The remaining counterterm, $R_{\mu\nu\rho\sigma}R^{\mu\nu\rho\sigma}$, gives us $C_3$. Again, this action only assumes that $\epsilon \ll 1$, and will therefore be applicable beyond slow-roll inflation.

Let us try to get the counterterms from another perspective. Since the counterterms only involve a pair of fields $h_{ij}$, and assuming that the rescaling of coordinates (dilatation)
\begin{equation} \label{eq:dilatation sym}
	a \to \lambda\, a\,,\quad \dd \tau \to \lambda^{-1} \dd \tau\,,\quad \dd x \to \lambda^{-1} \, \dd x\,,
\end{equation}
is a symmetry of the action, the kind of counterterms we expect to obtain are, schematically, of the form
\begin{equation}
	S\sim \int \dd \tau \dd^3\vb{x} \, a^{4-n-m}\: \tilde{B}\:  h_{ij}\left( \partial_\tau\right)^n \left(\partial_i \right)^m h_{ij}\,, 
\end{equation}
where $\tilde{B}$ is a constant that has energy dimensions of $[\tilde{B}] = 4-n-m$. Moreover, in addition to $M_P$ (whose dependence is already encoded in the tensor mode functions) the only physical scale --i.e.\ invariant under the dilatation symmetry-- that can give the correct dimensions to $\tilde{B}$ is $H$; i.e.\	 $\tilde{B} = B\times H^{4-n-m}$, where $B$ is a dimensionless constant. We further note that, since $H$ contains derivatives of the scale factor, the power of $H$ must always be positive since otherwise it would imply that we have derivatives in the denominator coming from non-local interactions. Therefore, the action of the counterterms has to be
\begin{align} \label{eq:Cts no general cov}
	S_{\rm cts}  = \int \dd \tau \dd^3\vb{x}\, \bigg\{ B_1 a^4 H^4 (h_{ij})^2 + a^2 H^2\left(B_2  (h'_{ij})^2 + B_3  (\partial_k  h_{ij})^2\right) +B_4 (\partial^2h_{ij})^2 +B_5 (\partial_k h'_{ij})^2  \bigg\} \,,
\end{align}
where $B_i$ are dimensionless constants. 
Again, we have used the free equation of motion of the modes $h_{ij}$ and (spatial) integration by parts to eliminate redundant terms. In addition, we have eliminated the temporal boundary terms that will not contribute to the tensor two-point correlation; see the discussion around eq.\  \eqref{eq:Not important BT}. Due to the dilatation symmetry, the constants $B_i$ can only renormalize quantities that are invariant under this symmetry. 
Eq.\ \eqref{eq:Cts no general cov} suggests that five different counterterms may be needed to renormalize the UV divergences we encounter computing the two-point function of tensor modes. However, with the previous procedure (in which we started from the covariant action) we only found three independent possible counterterms. The origin of the difference between both methods is the following: certain counterterms obtained by the second method --the use of the dilatation symmetry at the level of perturbations-- come from terms in the action that, before expanding in fluctuations, violate diffeomorphism invariance. Some terms do so in an explicit way, as is the case of the mass term $\sim h_{ij}h_{ij}$. However, other contributions are less obvious: although the last two terms of eq.\ (\ref{eq:Cts no general cov}) arise from a covariant action, they do not do so independently, but rather in such a way that $B_5 = -B_4$. 
This is why we must take special care in the search for the counterterms that renormalize certain observables in order not to have more free constants than {the maximum number that may be required a priori}.
Moreover, as we will see in Section \ref{sec:Tensor Ph in SR inf}, an incorrect use of counterterms can lead to erroneous predictions for the finite parts of the observable quantities at one loop.

For the calculation of the tensor power spectrum in dimensional regularization it is necessary to generalize the complete action --i.e.\ the action for the free fields, together with the interactions and counterterms-- to the case where the number of spatial dimensions is $d = 3+\delta$:
\begin{align} \label{eq:Action complete dim reg}
	\nonumber
	S = \int \dd \tau\, \mu^\delta\: \dd^{3 + \delta} \vb{x}\, \: a^{2+\delta} \Bigg\{ & \frac{M_P^2}{8} \left[  (h'_{ij})^2 - (\partial_k h_{ij})^2  \right]+\frac{1}{2} h_{ij} \partial_i\delta\phi \partial_j\delta\phi 
	-\frac{1}{4} h_{ik} h_{jk}\partial_i \delta\phi \partial_j \delta\phi  \\ & +\frac{1}{2} \left( \delta\phi ' \right) ^2- \frac{1}{2}(\partial_i\delta\phi)^2 -\frac{1}{2} a^2(\tau) \frac{\partial^2 V(\phi)}{\partial \phi^2}\eval_{\phi_0} \delta\phi^2  \nonumber \\
	& -  C_1 H^2 (h'_{ij})^2- C_2 H^2 (\partial_k h_{ij})^2  - \frac{C_3}{a^2} \left( (\partial_k h'_{ij})^2 - (\partial^2 h_{ij})^2 \right)  \Bigg\}
	\,.
\end{align}
For convenience, we introduce  the constant, physical (invariant under dilatations) energy scale $\mu$ to make the action dimensionless.\footnote{Although $\mu$ arises to compensate the dimensions introduced by the comoving measure $\dd^\delta \vb{x}$, we note that in $3+\delta$ spatial dimensions the only way to make $\mu^\delta \dd^\delta\vb{x}\: a^{\delta}$ invariant under dilatations, eq.\ \eqref{eq:dilatation sym}, is to impose $\mu$ to be a physical energy scale.} The factor $a^{\delta}$ comes from $\sqrt{-g}$, while all the other elements of the action remain unchanged in $3+\delta$ spatial dimensions.

\section{Tensor two-point correlation function at one loop} \label{sec:Tensor spectrum}

Once we have characterized the action that describes the coupling between the tensor and scalar sector, as well as the counterterms involved in the one-loop calculation of the two-point correlation of $h_{ij}$, we proceed to calculate the latter quantity. We will do so using the well-known in-in formalism (see e.g.\ \cite{Weinberg:2005vy, Chen:2010xka, Wang:2013zva} and the Appendix A of \cite{Ballesteros:2024zdp}).
Given a general Hermitian operator $\mathcal{O}(t)$ in the Heisenberg picture, its vacuum expectation value (at cosmic time $t$) is calculated as:
\begin{equation}
	\expval{\mathcal{O}(t)} = \frac{\bra{0} F^{-1}(t,-\infty_+) \mathcal{O}_I(t)F(t,-\infty_-) \ket{0}}{\bra{0} F^{-1}(t,-\infty_+) F(t,-\infty_-) \ket{0}} \,,
\end{equation}
where elements with a subscript $_I$ are in the interaction picture. The time evolution operator in this picture is defined by
\begin{equation}
	F(t,t_0) = T \exp \left( -i \int_{t_0}^t \dd t' H_I(t') \right) \quad {\rm and} \quad F^{-1}(t,t_0) = \overline{T} \exp \left(i \int_{t_0}^t \dd t' H_I(t') \right)\,,
\end{equation}
where ($\overline{T}$) $T$ denotes (anti)time-ordering.
The $i\omega$ prescription $-\infty_{\pm} = -\infty(1\pm i\omega)$, with an infinitesimal $\omega>0$ that eventually will be sent to 0, ensures that for $t \to -\infty$ the system projects onto the interaction picture vacuum $\ket{0}$.\footnote{This is motivated by the fact that at this time, all the modes of interest are well inside the horizon, being insensitive to the dynamics of the background and behaving freely, i.e.\ without interactions.}
Since $\mathcal{O}_I$ and the interaction Hamiltonian $H_I$ are in the interaction picture, the fields in these operators evolve according to the dynamics governed by the free Hamiltonian. For simplicity, we will omit the subscript $_I$ in what follows.
As we explicitly show in the Appendix \ref{ap: Bubbles and disconnected parts}, the vacuum expectation value of $\mathcal{O}(t)$ computed through the in-in formalism can be simplified by eliminating the bubble contributions,
\begin{equation}\label{in-in-f}
	\expval{\mathcal{O}(t)} = \bra{0} F^{-1}(t,-\infty_+) \mathcal{O}_I(t)F(t,-\infty_-) \ket{0}\eval_\textrm{no bubbles}\,.
\end{equation}
The operator giving the tensor two-point correlation is $\mathcal{O}(t) = h_{ij}(x) h_{ij}(y)$, where $x^0 = y^0 = t$.
Expanding eq.\ \eqref{in-in-f} in powers of the interaction Hamiltonian in the interaction picture, and organizing the expansion on the number of insertions of $H_I$ (indicated in the subscript) and working up to order $1/M_P^4$, we obtain
\begin{align}
	\expval{h_{ij}(\tau,\vb{x}) h_{ij}(\tau,\vb{y})}_0  & = 
	\begin{tikzpicture}
		\draw[decorate, decoration=snake] (0,0) -- (1,0);
	\end{tikzpicture}
	= \bra{0} h_{ij}(\tau,\vb{x}) h_{ij}(\tau,\vb{y})\ket{0} \,, \label{eq:2-Point 0}\\ 
	\expval{h_{ij}(\tau,\vb{x}) h_{ij}(\tau,\vb{y})}_1  &=
	\begin{tikzpicture}[baseline={(current bounding box.center)}]
		\draw[decorate, decoration=snake] (0,0) -- (0.5,0);
		\draw[decorate, decoration=snake] (0.5,0) -- (1,0);
		\fill[black] (0.5,0) circle (1.5pt);
		\draw (0.75,0.25) arc (360:0:0.25);
		\node[black] at (1.5,0.2) {$+$};
		\draw[decorate, decoration=snake] (2,0.2) -- (3,0.2);
		\draw[fill=white,cross] (2.5,0.2) circle (0.15);
	\end{tikzpicture} =2\Im{\int_{-\infty_-}^\tau\dd \tau'\bra{0} h_{ij}(\tau,\vb{x}) h_{ij}(\tau,\vb{y}) H_I(\tau') \ket{0}} \,,\label{eq:2-Point 1}\\
	\nonumber
	\expval{h_{ij}(\tau,\vb{x}) h_{ij}(\tau,\vb{y})}_2  & = 
	\begin{tikzpicture}[baseline={(current bounding box.center)}]
		\draw[decorate, decoration=snake] (6,0.5) -- (6.5,0.5);
		\draw[decorate, decoration=snake] (7,0.5) -- (7.5,0.5);
		\fill[black] (6.5,0.5) circle (1.5pt);
		\fill[black] (7,0.5) circle (1.5pt);
		\draw (7,0.5) arc (360:0:0.25);
		\node[black] at (8,0.5) {$+$};
		\draw[decorate, decoration=snake] (8.5,0) -- (8.5,0.5);
		\fill[black] (8.5,0.5) circle (1.5pt);
		\draw (8.75,0.75) arc (360:0:0.25);
		\draw[decorate, decoration=snake] (9.25,0) -- (9.25,0.5);
		\fill[black] (9.25,0.5) circle (1.5pt);
		\draw (9.5,0.75) arc (360:0:0.25);
	\end{tikzpicture}  = \int_{-\infty_-}^\tau\dd \tau' \int_{-\infty_+}^\tau\dd \tau'' \bra{0} H_I(\tau'') h_{ij}(\tau,\vb{x}) h_{ij}(\tau,\vb{y}) H_I(\tau')\ket{0}\quad \\ & \quad\quad\quad\quad- 2 \Re{ \int_{-\infty_-}^\tau \dd \tau' \int_{-\infty_-}^{\tau'} \dd \tau'' \bra{0} h_{ij}(\tau,\vb{x}) h_{ij}(\tau,\vb{y}) H_I(\tau') H_I(\tau'')\ket{0}} \,. \label{eq:2-Point 2}
\end{align}
Working in $d = 3+\delta$ spatial dimensions, the complete $H_I$ (considering both the interaction terms and the counterterms appearing in eq.\ \eqref{eq:Action complete dim reg}) is
\begin{align}
	\nonumber
	H_I(\tau) = \int \dd^{3+\delta}\vb{x} \, \mu^\delta a^{2+\delta} \bigg\{&-\frac{1}{2} h_{ij} \partial_i \delta\phi \partial_j \delta\phi + \frac{1}{4} h_{il} h_{lj} \partial_i \delta\phi \partial_j \delta\phi \\ & + C_1 H^2 (h'_{ij})^2 + C_2 H^2 (\partial_k h_{ij})^2
	+\frac{C_3}{a^2} \left((\partial_k h'_{ij})^2  - 	 (\partial^2 h_{ij})^2 \right) \bigg\} \,.
\end{align}
The momentum structure of the (connected) two-point correlation is, at all orders, 
\begin{equation}\label{eq:Def Power Spectrum}
	\expval{h_{ij}(\tau,\vb{x}) h_{ij}(\tau,\vb{y})} = \int \frac{\dd^{3+\delta}\vb{k}}{(2\pi)^3} e^{i\vb{k}\cdot (\vb{x} - \vb{y})} \frac{2\pi^2}{k^{3+\delta}} \mathcal{P}_h(\tau,k) \times 2\,,
\end{equation}
where the factor $\times 2$ is introduced so that $\mathcal{P}_h$ is the dimensionless power spectrum of each of the two polarizations of $h_{ij}$ in $3$ spatial dimensions. As discussed in Section \ref{sec:Constants in dim reg}, we set {$\delta = 0$} in all the constants, {which amounts to a choice of renormalization scheme.}

We now need to describe the fields in the interaction picture. The scalar field can be decomposed as
\begin{equation}\label{eq:Scalar Field def}
	\delta\phi(\tau,\vb{x}) = \int \frac{\dd^{3+\delta}\vb{k}}{(2\pi)^{3/2}} e^{i\vb{k}\cdot\vb{x}}\, \delta\phi_{\vb{k}}(\tau) \quad {\rm where} \quad  \delta\phi_{\vb{k}}(\tau) = \delta\phi_k(\tau) a_{\vb{k}} + \delta\phi_k^*(\tau) a^\dagger_{-\vb{k}}\,, 
\end{equation}
and the creation and annihilation operators satisfy the usual commutation rules. The function $\delta\phi_k(\tau)$ satisfies the free equation of motion  for the  fluctuations $\delta\phi$ in $d$ spatial dimensions and the Bunch-Davies initial conditions after a proper normalization \cite{Bunch:1978yq}. The tensor sector is more complicated because we need to take into account the polarizations of $h_{ij}$,
\begin{equation}
	h_{ij} (\tau,\vb{x}) = \int \frac{\dd^{3+\delta}\vb{k}}{(2\pi)^{3/2}} e^{i\vb{k}\cdot\vb{x}} \, h_{\vb{k},ij}(\tau)= \int \frac{\dd^{3+\delta}\vb{k}}{(2\pi)^{3/2}} e^{i\vb{k}\cdot\vb{x}} \sum_{\gamma} e^\gamma_{ij}(\vb{k}) h_{\vb{k}}^\gamma(\tau) \,.
\end{equation}
In $d$ spatial dimensions, the symmetric tensor $h_{ij}$ has $d\, (d+1)/2$ degrees of freedom. Imposing that it is traceless ($1$ constraint) and transverse ($d$ constraints), we finally have $(d+1)(d-2)/2$ physical degrees of freedom. Choosing the orthonormal basis $\{\hat{\vb{k}}, \vb{n}^1(\vb{k}), \dots , \vb{n}^{d-1}(\vb{k})\}$ for each {$\vb{k} \equiv k\, \hat{\vb{k}}$}, we can decompose $h_{ij}$ into the polarization basis
\begin{align}
	&(e^+) ^{\gamma_m}_{ij} = \frac{\vb{n}^1_i \vb{n}^1_j +\vb{n}^{2}_i \vb{n}^{2}_j +\dots + \vb{n}^{m}_i \vb{n}^{m}_j-m \, \vb{n}^{m+1}_i \vb{n}^{m+1}_j} {\sqrt{m+m^2}} \quad {\rm where} \quad m = 1,\dots,(d-2)\quad {\rm and}\\
	& (e^\times)^{\gamma_{ab}}_{ij} = \frac{1}{\sqrt{2}}\left(  \vb{n}^a_{i}\vb{n}^b_{j} + \vb{n}^b_{i}\vb{n}^a_{j}\right) \quad {\rm with} \quad a = 1,\dots,(d-2) \quad {\rm and} \quad b = (a+1),\dots,(d-1) \,.  
\end{align}
We omit the dependence on $\vb{k}$ for simplicity. We can use the freedom we have in choosing the polarization basis such that $e^\gamma_{ij}(\vb{k}) = e^\gamma_{ij}(-\vb{k})$, for convenience. 
There are $d-2$ polarizations in the sector $(+)$ and $(d-2)(d-1)/2$ in the sector $(\times)$, recovering the total number of degrees of freedom in $h_{ij}$. 
We note that this basis is properly orthonormalized, $e^{\gamma}_{ij} e^{\gamma'}_{ij} = \delta^{\gamma \gamma'}$.
Using the relation \cite{Cardoso:2002pa}
\begin{equation}
	\sum_\gamma e^\gamma_{ij} e^\gamma_{kl} = \frac{1}{2}\left(P_{ik}P_{jl} + P_{il}P_{jk} \right) -\frac{1}{d-1}P_{ij}P_{kl}\,, \quad \textrm{where we defined the projector} \quad P_{ij}\equiv  \delta_{ij} - \hat{\vb{k}}_i \hat{\vb{k}}_j\,, 
\end{equation}
we find that
\begin{equation}
	\sum_\gamma e^\gamma_{il} e^\gamma_{lj} \vb{p}_i \vb{p}_j = \frac{(d-2)(d+1)}{2(d-1)} p^2\sin^2\theta_{\vb{p},\vb{k}} \quad {\rm and} \quad \sum_\gamma \left( e^\gamma_{ij} \vb{p}_i \vb{p}_j\right)^2 = \frac{d-2}{d-1}\left( p^2\sin^2\theta_{\vb{p},\vb{k}}\right) ^2\,,
\end{equation}
being $\theta_{\vb{p},\vb{k}}$ the angle between $\vb{p}$ and $\vb{k}$. 
These are the only two relations {involving the polarization basis that} we will need for the calculation of the two-point correlation function.
We note that the dependence on $p$ and $\theta$ of these relations does not change when the number of spatial dimensions is modified. The dependence on $d$ is only reflected in a constant factor, which we will again take in $3$ spatial dimensions ($\delta =0$).

The tensor modes are decomposed into 
\begin{equation}
	h_{\vb{k}}^\gamma (\tau) = h_k(\tau) a^\gamma_{\vb{k}} + h_k^*(\tau) a^{\gamma,\dagger}_{-\vb{k}}\,,
\end{equation}
where again $h_k(\tau)$ satisfies the free equations of motion in $d$ spatial dimensions and has Bunch-Davies initial conditions after canonical normalization. The only non-zero commutators are 
\begin{equation} 
	\comm{a_{\vb{k}}}{a^\dagger_{\vb{p}}} = \delta(\vb{k} - \vb{p}) \quad {\rm and} \quad \comm{a^\gamma_{\vb{k}}}{a^{\gamma',\dagger}_{\vb{p}}} = \delta^{\gamma \, \gamma'} \delta(\vb{k} - \vb{p})\,.
\end{equation}

Now we report (in this order) the tree-level (tl) contribution, the quartic (1l,q) and cubic (1l,c) one-loop contributions, and the counterterms (cts) contribution, computed using eqs.\ \eqref{eq:2-Point 0}--\eqref{eq:2-Point 2}, to the power spectrum defined according to eq.\ \eqref{eq:Def Power Spectrum}: 
\begin{align}
	\mathcal{P}_h^{\rm tl}(\tau,k) & = \frac{k^{3+\delta}}{2 \pi^2} \abs{h_k(\tau)}^2\,, \label{eq:DimReg tl Ps}\\	
	\mathcal{P}_h^{\rm 1l,q}(\tau,k) & = \frac{k^{3+\delta}\mu^\delta}{4\pi^2} \int  \frac{\dd^{3+\delta} \vb{p}}{(2\pi)^3}  p^2 \sin^2\theta  \Im{h_{k}^2(\tau)  \int_{-\infty_-}^\tau \dd \tau' \, a^{2+\delta}(\tau')  h_{k}^{*2}(\tau') \abs{\delta\phi_{p}(\tau')}^2}\,, \label{eq:DimReg q Ps}\\ 
	\nonumber
	\mathcal{P}_h^{\rm 1l,c}(\tau,k) & = \frac{k^{3+\delta}\mu^{2\delta}}{4\pi^2} \int\frac{\dd^{3+\delta}\vb{p}}{(2\pi)^3}\frac{1}{2} p^4 \sin^4\theta\, \times \\ \nonumber
	& \hspace{-1.7cm} \times \Bigg[ \abs{h_k(\tau)}^2  \int_{-\infty_-}^\tau \dd \tau' \: a^{2+\delta}(\tau') h^*_k(\tau') \int_{-\infty_+}^\tau \dd \tau'' \: a^{2+\delta}(\tau'') h_k(\tau'') \delta\phi_p(\tau'') \delta\phi_p^*(\tau') \delta\phi_q(\tau'') \delta\phi_q^*(\tau') \\
	  & \hspace{-1.7cm}   -2 \Re \Bigg\{ h_k^2(\tau) \int_{-\infty_-}^\tau \dd \tau' \: a^{2+\delta}(\tau') h^*_k(\tau') \int_{-\infty_-}^{\tau'} \dd \tau'' \: a^{2+\delta}(\tau'') h_k^*(\tau'')  \delta\phi_p(\tau') \delta\phi_p^*(\tau'') \delta\phi_q(\tau') \delta\phi_q^*(\tau'')\Bigg]\Bigg\}\,, \label{eq:DimReg c Ps} \\ 
	 \nonumber
	\mathcal{P}_h^{\rm cts}(\tau,k) & = \frac{k^{3+\delta}\mu^\delta}{4\pi^2} 8 \Im \Bigg\{ h^2_k(\tau) \int_{-\infty_-}^\tau\dd \tau' \: a^{2+\delta}(\tau') \bigg[C_1H^2  \left(\frac{d\, h_k^*(\tau')}{d\tau'}\right)^2 + C_2 H^2k^2 h_k^{*2}(\tau') \\
	 &     \quad \quad\quad\quad\quad\quad\quad  \quad \quad\quad\quad\quad\quad\quad  \quad \quad\quad\quad+ C_3\frac{k^2}{a^2(\tau')} \bigg(\left(\frac{d\, h_k^*(\tau')}{d\tau'}\right)^2 - k^2 h_k^{*2}(\tau') \bigg) \bigg] \Bigg\}\,, \label{eq:DimReg cts Ps}
\end{align} 
where here and in what follows, $q = \abs{\vb{k} - \vb{p}}$ is the modulus of the transfer momentum (being $\vb{k}$ the external momentum and $\vb{p}$ the internal one). We emphasize that this result holds for any single field inflationary model as long as $\epsilon\ll 1$, beyond slow-roll inflation.

We recall that we choose a scheme where in all the constants we take $\delta = 0$. We evolve all the modes appearing in each of the contributions to $\mathcal{P}_h$ in $3+\delta$ spatial dimensions, including the external modes with comoving momentum $k$. However, we note that in \cite{Senatore:2009cf} (see also \cite{delRio:2018vrj}) the physics considered as IR was evaluated in $3$ spatial dimensions, taking $\delta = 0$ in $k^{3+\delta}$ and also in the external modes; but the $k$-dependent modes that come from the interaction vertices --e.g.\ in our case the modes $h_k(\tau')$ and $h_k(\tau'')$ in the expressions above-- were allowed to evolve in $3+\delta$ spatial dimensions in those references. Such a choice can lead to incorrect results for the finite part of the correlation functions. We will come back to this point later. 

{Given} that $\mathcal{P}_h^{\rm tl}$ does not involve a $\sim 1/\delta$ pole, we can safely take $\delta = 0$ in eq.\ \eqref{eq:DimReg tl Ps}. However, since the constants of the counterterms will take care of absorbing the $1/\delta$ divergences of the one-loop contribution, we must keep the dependence on $\delta$ of the elements composing $\mathcal{P}_h^{\rm cts}$.

To perform the loop integrals, a useful change of variable is to consider the transfer momentum $q = \abs{\vb{k} - \vb{p}}$  as the new integration variable instead of $\theta$ \cite{Espinosa:2018eve}. The volume element will be:
\begin{equation}
	\dd^{3+\delta}\vb{p} = p^{2+\delta} \left( \sin \theta \right)^{1+\delta} \dd p\: \dd \theta \: \dd \phi = p^\delta \sin^\delta\theta\: \frac{p\: q}{k}\: \dd p\: \dd q \: \dd \phi\,,
\end{equation}
where $q\in\{\abs{k-p},\: k+p\}$ and $\sin^2 \theta = -{(k^2-(p+q)^2) (k^2-(p-q)^2)}/({4 k^2p^2})$.
To evaluate the integrals systematically, we first expand the integrand at first order in $\delta$, without expanding the term $p^\delta$.\footnote{By expanding $\sin^\delta\theta$ around $\delta = 0$, at first order we will obtain $\delta\:\log \sin\theta$ that naively seems to contain a $\log p$ term. Although, in general, logarithms can be problematic, in this case they are not, since the sine function is bounded, and therefore even if we obtain logarithms they are not UV divergent.} We can first analytically perform the time integrals, taking special care when evaluating them in the lower limit to properly include the $i\omega$ prescription, that ensures their convergence. Then we compute the momentum integral in $p$ following the procedure described in Section \ref{sec:dimregprocedure}.

\subsection{Tadpoles}

The tadpoles that appear in eq.\ \eqref{eq:2-Point 2} do not contribute to the power spectrum due to their momentum structure --they are disconnected--, but they do contribute to the two-point correlation and have to be taken into account.
Using the property that the disconnected parts of the $n$-point correlations are constructed in terms of connected correlations of lower orders (see Appendix \ref{ap: Bubbles and disconnected parts}):
\begin{equation}
	\expval{h_{ij}(\tau,\vb{x})h_{ij}(\tau,\vb{y})} \supset
	\begin{tikzpicture}[baseline={(current bounding box.center)}]
		\draw[decorate, decoration=snake] (8.5,0) -- (8.5,0.5);
		\fill[black] (8.5,0.5) circle (1.5pt);
		\draw (8.75,0.75) arc (360:0:0.25);
		\draw[decorate, decoration=snake] (9.25,0) -- (9.25,0.5);
		\fill[black] (9.25,0.5) circle (1.5pt);
		\draw (9.5,0.75) arc (360:0:0.25);
	\end{tikzpicture}
	= \expval{h_{ij}(\tau,\vb{x})} \expval{h_{ij}(\tau,\vb{y})}\,.
\end{equation}
We will now show that this contribution vanishes in a model-independent way. Therefore, we can work now in $3$ spatial dimensions, as no regularization is required.\footnote{Since this calculation gives zero, working in $3+\delta$ dimensions we would obtain the same result.} We have
\begin{equation}
	\expval{h_{ij}(\tau,\vb{x})} = - \Im{\int_{-\infty_-}^\tau \dd \tau'\: a^2(\tau') \int \frac{\dd^3\vb{p}}{(2\pi)^3} \frac{\dd^3\vb{q}}{(2\pi)^3} \delta(\vb{p}) \sum_\gamma e^\gamma_{ij}(\vb{p}) e^\gamma_{kl}(\vb{p}) q_k q_l h_p(\tau) h_p^*(\tau') \abs{\delta\phi_q(\tau')}^2}\,.
\end{equation}
To calculate the integral in $\vb{q}$, defining $q_p = \vb{q}\cdot\hat{\vb{p}}$, $q_1 = \vb{q}\cdot \vb{n}^1(\vb{p})$ and $q_2 = \vb{q}\cdot \vb{n}^2(\vb{p})$, we need to analyze
\begin{equation}
	\sum_\gamma e^\gamma_{ij}(\vb{p}) e^\gamma_{kl}(\vb{p}) q_k q_l = \frac{1}{\sqrt{2}}\left( e^+_{ij}(\vb{p})\left(q_1^2 - q_2^2 \right) +2\,e^\times_{ij}(\vb{p}) q_1 q_2  \right) \,,
\end{equation}
so we have to calculate
\begin{equation}
	\int \dd q_p \dd q_1 \dd q_2 \left(  e^+_{ij}(\vb{p})\left(q_1^2 - q_2^2 \right) +2\,e^\times_{ij}(\vb{p}) q_1 q_2  \right) \abs{\delta\phi_q(t')}^2\,.
\end{equation}
We note that $\delta \phi_q$ only depends on $q = \sqrt{q_p^2 + q_1^2 + q_2^2}$ and therefore is symmetric under $q_1 \leftrightarrow q_2$ and is even in the variables $q_1$ and $q_2$, so that the integral in $\vb{q}$ vanishes. We conclude that the one-point correlation of the tensor modes is identically zero,
\begin{equation}
	\expval{h_{ij}(\tau,\vb{x})} = 
	\begin{tikzpicture}[baseline={(current bounding box.center)}]
		\draw[decorate, decoration=snake] (8.5,0) -- (8.5,0.5);
		\fill[black] (8.5,0.5) circle (1.5pt);
		\draw (8.75,0.75) arc (360:0:0.25);
	\end{tikzpicture} 
	= 0\,,
\end{equation}
as expected for the vacuum expectation value of a non scalar fluctuation in a SO(3) invariant theory, since otherwise the vacuum would not be homogeneous and isotropic.

\subsection{Tensor power spectrum in slow-roll inflation at one loop} \label{sec:Tensor Ph in SR inf}

In this section we finally calculate the tensor power spectrum in the case of slow-roll inflation at zeroth order in $\epsilon$. To do so, the first step is to calculate the free dynamics of the scalar, $\delta\phi_k(\tau)$, and tensor, $h_k(\tau)$, modes in $d$ spatial dimensions. We start by analyzing the scalar fluctuation.
Since we have introduced the factor $\mu$ into the action \eqref{eq:Action complete dim reg}, the field $\delta \phi (x)$ defined in eq.\ \eqref{eq:Scalar Field def} has dimensions of energy $\left[ \delta \phi (x) \right] = 1$ and then $\left[ \delta \phi_{\vb{k}}(\tau) \right] = -2-\delta$. The commutation rules are  $\comm{a_{\vb{k}}}{a_{\vb{p}}^\dagger} = \delta(\vb{k}-\vb{p})$, and then $\left[ a_{\vb{k}} \right] = -3/2 - \delta/2$ and $\left[\delta\phi_k(\tau) \right] = -1/2 - \delta/2 $. 
Taking into account this dimensional analysis, the initial conditions of the modes must include the physical scale $\mu$ to compensate for the change of the dimension of the mode in $3+\delta$ with respect to the case $\delta =0$. These initial conditions arise naturally by including in the usual Bunch-Davies initial conditions the factor that canonically normalizes the field $\delta\phi$. Writing both the equations of motion and the initial conditions, we have \cite{Senatore:2009cf}:\footnote{In \cite{Senatore:2009cf}, the effects of modes living in $3+\delta$ spatial dimensions were included consistently. Some earlier analyses using dimensional regularization did not properly take this effect into account (e.g.\ \cite{Weinberg:2005vy,Chaicherdsakul:2006ui}), leading to incorrect results that violated the dilatation symmetry.}
\begin{equation}
	\delta\phi_k'' + (2+\delta)aH \: \delta \phi'_k + k^2 \delta\phi_k = 0  \quad {\rm and} \quad \delta\phi_k(\tau) \xrightarrow{k\tau \to- \infty} \frac{1}{a^{1+\delta/2}(\tau) \mu^{\delta/2}}\frac{e^{-ik\tau}}{\sqrt{2k}}\,.
\end{equation}
{Although for the purpose we are interested in (i.e.\ obtaining the one-loop power spectrum of tensor modes at zeroth order in $\epsilon$)  this differential problem can be solved exactly, in more complicated situations
this may not be the case. Therefore, we solve these equations in a way that can be generalized for more complicated computations. We can make the expansion:
\begin{equation}
	\delta\phi_k(\tau) = \delta\phi_k^{(0)}(\tau) + \delta\: \delta\phi_k^{(1)}(\tau) + \order{\delta^2}\,,
\end{equation}
where we cut the expansion to the linear order because, as we have already seen, the following corrections will not contribute to the power spectrum at one loop once we take the limit $\delta \to 0$.\footnote{As explained, this is only true 
{if, by making the expansion around $\delta = 0$, no terms of type $\log k$ arise.}
By analyzing the equation of motion of the modes, together with their initial conditions --intimately related to their UV ($k\to \infty$) behavior-- we notice that the only logarithms that can arise in the expansion around $\delta = 0$ are temporal, $\sim \log (-\tau)$.} The equations of motion (together with the asymptotic boundary condition) that we have to solve are
\begin{align}
	\left( \delta\phi_k^{(0)}\right)  '' + 2aH \: \left( \delta \phi_k^{(0)}\right) ' + k^2 \delta\phi_k^{(0)} = 0 \,, & \quad \delta\phi_k^{(0)}(\tau) \xrightarrow{k\tau \to- \infty} \frac{1}{a(\tau)}\frac{e^{-ik\tau}}{\sqrt{2k}}\,,\\
	\left( \delta\phi_k^{(1)}\right) '' + 2aH \: \left( \delta \phi_k^{(1)}\right) ' + k^2 \delta\phi_k^{(1)} = -aH \: \delta\phi_k^{(0)}  \,, & \quad \delta\phi_k^{(1)}(\tau) \xrightarrow{k\tau \to- \infty} -\frac{1}{2}\log \left(\mu a(\tau) \right) \frac{1}{a(\tau)}\frac{e^{-ik\tau}}{\sqrt{2k}}  \,.
\end{align}
Using that $a(\tau) H = -1/\tau$ at zeroth order in $\epsilon$, the result is
\begin{align} \nonumber
	\delta\phi_k (\tau) & = i \frac{H}{\sqrt{2 k^3}}e^{- i k \tau} (1 + i k \tau) \times \\& \quad \times \bigg\{1+ \frac{\delta}{2}\bigg[  \frac{(1-i k \tau) }{(1+i k \tau )}e^{2 i k \tau }(i\pi - \text{Ei}(-2 i k \tau )) +  \frac{2}{1+ik \tau}+ \log \left( -\frac{H \tau }{\mu} \right) \bigg]  + \order{\delta^2}\bigg\} \,, \label{eq:Mode evolution} 
\end{align}
where ${\rm Ei}(z)=-\int_{-z}^\infty \dd x \, e^{-x}/x$ is the exponential integral function.\footnote{In previous studies (see for example \cite{Senatore:2009cf, delRio:2018vrj, Comelli:2022ikb}), only the one-loop logarithmic contribution to observable quantities (e.g.\ the power spectrum) was studied. For this purpose, only the term $\sim \delta \log (-H\tau / \mu)$ in the time evolution of the modes in $3+\delta$ spatial dimensions (eq.\ \refeq{eq:Mode evolution}) was considered. As we shall see, this is not enough to obtain the complete one-loop contribution to the observable quantities: it is necessary to include all the terms appearing in eq.\ \refeq{eq:Mode evolution}. This highlights the motivation to develop a procedure, such as the one introduced in Section \ref{sec:dimregprocedure}, that allows to calculate loop integrals completely in dimensional regularization in the context of inflation.}
As expected, in eq.\ \refeq{eq:Mode evolution} no terms arise that worsen the convergence of the integrals in dimensional regularization, such as the logarithmic UV behavior $\log k$; see the discussion around eq.\ \eqref{eq:log div expanded}.
The calculation of the free evolution of the tensor modes in $d$ spatial dimensions is analogous to the scalar ones, and leads to an identical time evolution except for a normalization factor $h_k(\tau) = 2/M_P\, \delta\phi_k(\tau)$, as long as we analyze a slow-roll dynamics at zeroth order in $\epsilon$.

Once we know the dynamics of the free scalar and tensor modes, we can compute 
the various contributions to the tensor power spectrum described in eqs.\ \eqref{eq:DimReg tl Ps}--\eqref{eq:DimReg cts Ps}, following the methodology presented in Section \ref{sec:dimregprocedure} for the calculation of loop integrals in dimensional regularization:
\begin{align} \label{conts1}
	\mathcal{P}_h^{\rm tl}(\tau,k)
	& = \frac{H^2}{M_P^2 \pi^2} \,\left(1+(k\tau)^2 \right) \,,\\ \label{conts2}
	\mathcal{P}_h^{\rm 1l,q}(\tau,k)
	& = \frac{H^4}{M_P^4 \pi^4} \, \frac{1}{8} \Re{e^{2ik\tau}(i+k\tau)^2 \left(-i\pi+{\rm Ei}(-2ik\tau) \right)  } \,,\\ \nonumber
	\mathcal{P}_h^{\rm 1l,c}(\tau,k)
	& = \frac{H^4}{M_P^4 \pi^4} \, \bigg[ \frac{3}{80}\left( 1 + (k\tau)^2 \right) \left(\frac{1}{\delta} + 2 \log \frac{k}{\mu a(\tau)} \right) \\  & \quad\quad\quad\quad\quad- \frac{1}{20} \Re{e^{2ik\tau}(i+k\tau)^2 \left(-i\pi+{\rm Ei}(-2ik\tau) \right)  } \bigg] \,, \label{conts3} \\ \nonumber
	\mathcal{P}_h^{\rm cts}(\tau,k)
	& = \frac{H^4}{M_P^4 \pi^4} \, \Bigg[ c_1^{\rm fin} + c_2^{\rm fin}(k\tau)^2 + c_3^{\rm fin}(k\tau)^4 + \left(c_1^{\rm div} + c_2^{\rm div}(k\tau)^2 + c_3^{\rm div}(k\tau)^4 \right)\left( \frac{1}{\delta} + \log \frac{k}{\mu a(\tau)}\right) \\
	& \hspace{-1.2cm} + \Re{e^{2ik\tau}(i+k\tau)\left( i c_1^{\rm div} + c_1^{\rm div} (k\tau) -i \left(c_1^{\rm div}-c_2^{\rm div} \right)(k\tau)^2+c_3^{\rm div}(k\tau)^3  \right)\left(-i\pi+{\rm Ei}(-2ik\tau) \right)  }  \Bigg]\,, \label{cont4}
\end{align}
where, for convenience, we have defined $c_i$ as follows:
\begin{equation} \label{eq:Cts redef}
	C_1 = -\frac{c_1-3c_2+3c_3}{16\pi^2} \,,\quad  C_2 = -\frac{c_1+c_2-3c_3}{16\pi^2}\,, \quad C_3 = \frac{c_3}{8\pi^2} \,,
\end{equation}
with each $c_i$ split into a divergent and a finite part 
\begin{align}
c_i\equiv \frac{c_i^{\rm div}}{\delta} + c_i^{\rm fin}\,.
\end{align}

Each contribution to the tensor power spectrum, eqs.\ \eqref{conts1}--\eqref{cont4},  is invariant under the dilatation symmetry. The eqs.\ \eqref{conts1}--\eqref{cont4}, where we have redefined $c_i^{\rm fin}$ to absorb all the finite constants from the loop and counterterm contributions to the power spectrum,  are valid for any conformal time $\tau$ during slow-roll inflation. 

We are now going to analyze the most relevant aspects of the different contributions to the power spectrum. This analysis will make explicit the importance of the fact that both external and internal modes evolve in $3+\delta$ spatial dimensions and the relevance of the adequate choice of counterterms. 

Expanding the integrand of the quartic contribution --see eq.\ \eqref{eq:DimReg q Ps}-- at {zeroth} order
in $\delta$, we find that the corresponding contribution to the power spectrum is zero. At this order, the momentum dependence is purely polynomial over the whole range of values of $p$, i.e.\ both in the UV and in the IR, and then:
\begin{equation}
	\int_0^\infty \dd p \: p^{\delta + n} = \lim_{L\to\infty}\bigg(\int_0^L \dd p \: p^{n} - \frac{L^{n+1}}{n+1}\bigg) +\order{\delta} \xrightarrow{\delta \to 0} 0, \quad \forall n> 0\,. \footnote{This equation is also true $\forall n \neq -1$.}
\end{equation}
Since the UV behavior of the integrand of eq.\ \eqref{eq:DimReg q Ps} at this order has no $1/p$ terms, it has no logarithmic divergences (that would give rise to $1/\delta$ poles). At first order in $\delta$, $1/p$ terms appear, resulting in a finite non-zero contribution to the power spectrum, as seen in eq.\ \eqref{conts2}. However, expanding the integrand of the cubic contribution, eq.~\eqref{eq:DimReg c Ps}, at {zeroth}
order in $\delta$, we obtain logarithmic divergences that give rise to the only divergent term (a $1/\delta$ pole) of the one-loop power spectrum. As in the quartic case, the expansion of the integrand at first order in $\delta$ leads to a finite contribution, see  eq.\ \eqref{conts3}. We point that some these finite contributions would be missed if the spatial dimensionality of the modes was improperly chosen (e.g.\ by setting $\delta =0$ for the external modes). 

Let us now discuss the contribution of the counterterms, eq.\ \eqref{cont4}. An interesting aspect of it is that the following structure, which allows to absorb the one-loop divergences,
\begin{equation}\label{eq:P CTs simple}
	\mathcal{P}_h^{\rm cts}(\tau,k) \supset \frac{H^4}{M_P^4 \pi^4} \times \bigg[ s_1  + s_2 (k\tau)^2 + s_3 (k\tau)^4 \bigg] \,,\quad \textrm{with}\quad 
	s_i = s_i^{\rm fin}+ s^{\rm div}_i/\delta\,,
\end{equation}	
can appear in eq.\ \eqref{cont4} even if other counterterms, different from those coming from eq.\ \eqref{eq:Cts general cov}, are used. Indeed, if we use  the counterterms in eq.\ \eqref{eq:Cts general cov} --as we have done to obtain eq.\ \eqref{cont4}--, then $s_i = c_i$; but if we used the action \eqref{eq:Cts no general cov} for the counterterms, we could either take $B_4\neq 0$ and $B_5 = 0$, or take $B_4 = 0$ and $B_5 \neq 0$, and in both cases we would get the structure described by eq.\ \eqref{eq:P CTs simple}. However, this degeneracy breaks down when we look at the finite part of $\mathcal{P}_h^{\rm cts}$ generated by the different 
counterterms, see the second line of eq.\ \eqref{cont4} (which comes from the modes evolving in $3+\delta$ spatial dimensions).  Being finite, this contribution is of physical importance, which implies that one needs to choose the counterterms appropriately to get it right. An incorrect choice of counterterms may still allow to absorb the one-loop divergences, but leading to an erroneous prediction for the finite part of the total tensor power spectrum. This leads us to an interesting conclusion: in the calculation of the power spectrum induced by scalar modes, both the scalar-tensor interactions described in \eqref{eq:Action for the fluctuations} and the counterterm interactions coming from \eqref{eq:Cts action general} play an equally important role, which is natural from an effective field theory point of view.

Looking at eq.\ \eqref{eq:DimReg c Ps}, a noteworthy point is that, in $\mathcal{P}_h^{\rm 1l,c}$, the coefficient of the logarithmic contribution $\sim \log k$ is equal to two times the coefficient of the divergent term $\sim 1/\delta$. The reason why these two coefficients are not the same is the $\delta$ in the factor
$k^{3+\delta}$ coming from the volume element in the integral eq.~\eqref{eq:DimReg c Ps}, which introduces an extra $\log k$ with respect to the one that comes solely from the loop integral.\footnote{In \cite{Senatore:2009cf} a structure ${1}/{\delta} + \log {k}/{\mu} + \log{H}/{k}$ (without the factor 2) was obtained taking 3 spatial dimensions both for the external modes and the volume element, but this choice does not allow to get the complete (finite) spectrum at one loop.}

Let us now impose that the total power spectrum at one loop has to be finite in the limit $\delta\to0$. This requires the divergent part of the counterterms to be determined by the choice: 
\begin{equation} \label{eq:Cts sol}
	c_1^{\rm div} = c_2^{\rm div} = -\frac{3}{80} \quad {\rm and} \quad c_3^{\rm div} = 0 \,.
\end{equation}
This means that to renormalize the divergences we only need the counterterms $R^2$ and $R_{\mu\nu}R^{\mu\nu}$ (and therefore $\square R$ and $R_{\mu\nu\rho\sigma}R^{\mu\nu\rho\sigma}$ can be discarded). The final solution for the power spectrum is:
\begin{align} \nonumber 
\mathcal{P}_h(\tau,k) & = \frac{H^2}{M_P^2 \pi^2} \, \left(1+(k\tau)^2 \right) + \frac{3}{80} \left(  \frac{H^2}{M_P^2 \pi^2} \right) ^2\, \bigg[ \alpha + \beta (k\tau)^2+\gamma(k\tau)^4 + \left(1+(k\tau)^2 \right) \log\frac{k}{\mu a(\tau)}\\
	& \quad\quad\quad\quad\quad\quad\quad\quad\quad\quad\quad\quad + \Re{e^{2ik\tau}(i+k\tau)^2 \left(-i\pi+{\rm Ei}(-2ik\tau) \right)} \bigg] + \order{\frac{H^2}{M_P^2 \pi^2}}^3\,, \label{eq:tensorPS}
\end{align}
where $\alpha$, $\beta$ and $\gamma$ are free constants that should be determined using measurements of $\mathcal{P}_h(\tau,k)$.\footnote{Note that $\mu$ does not count as an extra constant, since it does not come from the counterterms. The effect of $\mu$ can be completely absorbed, in this case, by the constants $\alpha$ and $\beta$.}$^,$\footnote{In \cite{Campos:1994xx,Frob:2012ui} a similar result is obtained for the tensor spectrum induced by scalar modes conformally coupled to gravity.
In that case, a conformal transformation allows to rewrite the problem as an equivalent one with a flat background metric, where the loop integrals in dimensional regularization are known. We note that starting from a conformal scalar coupling and a flat potential (in practice a cosmological constant), switching to the Einstein frame where the coupling to gravity is minimal --as in our case--, slow-roll inflation is not recovered \cite{Kofman:2007tr}.} The factor 2 that accompanied the term $\sim\log k$ in the cubic contribution, eq.\ \eqref{conts3}, turns to 1 once the effect of the counterterms is taken into account. Evaluating eq.\ \eqref{eq:tensorPS} at the end of inflation, i.e.\ taking $\tau \to 0$, we get:
\begin{align}
\label{eq:latetimePS}
	\lim_{\tau\to 0^{-}}\mathcal{P}_h(\tau,k) = \frac{H^2}{M_P^2 \pi^2}  + \frac{3}{80}\left(  \frac{H^2}{M_P^2 \pi^2} \right) ^2 \left(  \alpha -\gamma_E + \log \frac{H}{2\mu}\right) + \order{\frac{H^2}{M_P^2 \pi^2}}^3\,,
\end{align}
where $\gamma_E$ is Euler-Mascheroni's constant. Although eq.\ \eqref{eq:tensorPS}  shows a behavior that seems to give late-time divergence $\sim \log a(\tau)$ \cite{Chen:2016nrs}, when taking the limit $\tau\to0$ this apparent divergence is shown to be a mirage, due to a cancellation with the $\textrm{Ei}(-2ik\tau)$ term. This cancellation is very subtle: to be realized it requires a cancellation among the finite parts of the quartic and cubic contributions, whereas the counterterms give a contribution that is finite at late times.\footnote{We thank M.\ Braglia and L.\ Pinol for a discussion about this point.}
We stress that the cancellation of the late-time divergence happens thanks to the external modes and the measure of integration being in $3+\delta$ spatial dimensions in dimensional regularization.

The cancellation of the late-time divergence does not happen by chance. It is associated to the (zero) mass of the tensor modes of the metric being protected (see \cite{Pimentel:2012tw,Assassi:2012et,Senatore:2012ya} for analogous arguments on the scalar variable $\zeta$).\footnote{The variable $h_{ij}$ defined according to the decomposition of the metric eq.\ \eqref{eq:SVT_Decomposition} classically freezes at late times. Using a different decomposition of the metric, tensor modes do not necessarily freeze. Similarly, the statement that at the quantum level the limit $\tau\to0$ of $\mathcal{P}_h$ is constant in time depends on the metric decomposition.} 
Failure to cancel the late-time divergence would signal the generation of a loop-level induced mass. In general, this would imply the breakdown of perturbation theory in the late-time limit $\tau\to 0$. This phenomenon has been studied in the literature (see e.g.\ \cite{Burgess:2009bs,Gorbenko:2019rza}), with non-perturbative methods developed to handle such divergences.

The cancellation of the late-time divergence of $\mathcal{P}_h$, together with the dilatation symmetry, implies that the result cannot depend on the momentum $k$ at late times, since that is the only comoving energy scale of the problem. This is why the loop level contribution to the tensor power spectrum exhibits no running in the limit $\tau \to 0$, yielding a result that is completely scale invariant. {This scale invariance is softly broken by higher-order slow-roll corrections. More in general, if additional comoving scales $k_*$ are present, e.g.\ in transient ultra slow-roll inflation,}
despite the cancellation of the late-time divergences, the one-loop contribution to $\mathcal{P}_h$ may depend on the momentum $k$, having a non-trivial running and also being dilatation symmetric.

Before closing this section let us comment on the size of the loop corrections to the tensor power spectrum in slow-roll inflation. Assuming e.g.\ $\epsilon\sim 10^{-3}$ and noting that $H^2 / M_P^2 \sim \epsilon\, \mathcal{P}_\zeta \sim 10^{-3} \times 10^{-9} = 10^{-12}$, we expect $\mathcal{P}_h^{\rm 1l} / \mathcal{P}_h^{\rm tl} \sim \alpha \times 10^{-12}$, see eq. \eqref{eq:latetimePS}. For $\alpha \sim \order{1}$ the late-time one-loop correction to the tree-level power spectrum amounts to a negligible shift with respect to the tree-level prediction. 

The significance of our analysis  lies not in predicting the one-loop correction to the tensor power spectrum for this specific case but in developing a method that enables systematic calculation of the finite parts of loop integrals in dimensional regularization. Although the relatively simple example we have used to illustrate the method does not offer a quantitatively relevant phenomenological result, the method is applicable across various scenarios where loop corrections might be numerically important. 

\section{Regularization with a cutoff} \label{sec:Cutoff}

In this section we recalculate the one-loop tensor power spectrum induced by scalar fluctuations during slow-roll inflation, this time using a cutoff as a regulator. As we mentioned earlier, a key principle of renormalization is that modifications of the UV physics that allow to regularize should not be observable, so the result should be independent of the regulator once we take into account the effect of the counterterms. It is also well known that certain regulators are more convenient than others. In particular, a regulator that breaks a given symmetry will produce results that reflect this breaking (see e.g.\ \cite{Negro:2024bbf} for a recent discussion in the context of inflation).
Although a priori using a symmetry-breaking cutoff is not a problem per se, it can be inconvenient to use this type of regulator instead of others in which the symmetries of the system are respected. In the case we have used to illustrate our dimensional regularization method, we find that although the use of a cutoff as a regulator breaks diffeomorphism invariance, we recover the correct power spectrum thanks to the use of appropriate counterterms.

Since the use of any cutoff breaks diffeomorphism invariance, the counterterms capable of renormalizing the tensor power spectrum in the example we have explored  cannot be those derived from eq.\ \eqref{eq:Cts action general}.
Instead, we can obtain the necessary counterterms for renormalizing with a cutoff by means of the arguments developed around eq.\ (\ref{eq:Cts no general cov}). 
To arrive at that expression we used that the counterterms must be dilatation invariant (and assumed that the coefficients $B_i$ are also dilatation invariant). Regarding the nature of the cutoff, it is then necessary to distinguish between a cutoff associated with a physical or a comoving energy scale: the former is invariant under the dilatation symmetry, eq.\ \eqref{eq:dilatation sym}, while the latter is not. If the cutoff is chosen to be comoving, the coefficients $B_i$ must absorb (divergent) comoving quantities that are not dilatation invariant, violating the original assumption on $B_i$.} This means that the possible counterterms allowed with a comoving cutoff are not just those in eq.\ \eqref{eq:Cts no general cov}. In contrast, using a physical, dilatation-invariant cutoff, the full counterterm action is described by eq.\ \eqref{eq:Cts no general cov}. That is why we use a physical cutoff, $\Lambda$, such that
\begin{align} \label{cutt2}
p_{\rm phys} = \frac{p}{a} <\Lambda\,,
\end{align} 
where $p$ is the modulus of the comoving momentum running in the loop.

After these considerations, we can calculate the different contributions to the tensor power spectrum starting with the counterterms, simply applying the in-in formalism to eq.\ \eqref{eq:Cts no general cov}:
\begin{equation} \label{eq:tensorPS cutoff}
	\mathcal{P}_h^{\rm cts}(\tau,k) =\frac{3}{80} \frac{H^4}{M_P^4 \pi^4} \left( \alpha + \beta (k\tau)^2 + \gamma (k\tau)^4 + \kappa\Re{e^{2ik\tau} \left( i+k\tau \right) ^2 \left( -i\pi + \text{Ei}(-2ik\tau)\right) }  \right) \,.
\end{equation}
We have four free constants ($\alpha$, $\beta$, $\gamma$ and $\kappa$) to determine with the experiment --after absorbing the divergences of the one-loop contribution--, whereas in dimensional regularization (see eq.\ \eqref{cont4}) we had only three. The additional constant ($\kappa$) comes from the mass counterterm that we have included to regularize with a cutoff but not in dimensional regularization, since it violates diffeomorphism invariance. As we are going to see, including this counterterm with cutoff regularization is crucial to make the results compatible with dimensional regularization. 

To calculate each one-loop contribution to the tensor power spectrum, since the limits of integration of the momentum integrals depend on time through the cutoff, eq.\ \eqref{cutt2}, we will first do the momentum integral, unlike in dimensional regularization. Considering first the quartic contribution, to regularize the momentum integral --consider eq.~\eqref{eq:DimReg q Ps} with $\delta = 0$, taking care of the integration order-- we impose a cutoff $a(\tau') \Lambda$ on the comoving momentum \cite{Senatore:2009cf}, which makes this integral finite and regulator-dependent. The subsequent time integral is directly finite, so no temporal cutoff is required, contrary to what happens with the cubic contribution (as we will discuss later). We obtain:
\begin{equation}
	\mathcal{P}_h^{\rm 1l,q}(\tau,k) =  - \frac{\Lambda^2 \left(2H^2 + \Lambda^2 \right) }{36\pi^4 M_P^4} \left( 2 + \Re{e^{2ik\tau} \left( i+k\tau \right) ^2 \left( -i\pi + \text{Ei}(-2ik\tau)\right) } \right) \,.
\end{equation}
One might wonder why we choose the integration time $\tau'$ when imposing the cutoff $a(\tau')\Lambda$ in the momentum integral, instead of employing another time e.g.\ the external one $\tau$. The choice of $\tau'$ is the most natural prescription; after all, we are regularizing a loop integral that arises from inserting one $H_I(\tau')$. Moreover, selecting a time such as $\tau$ in the regulator would introduce divergences that cannot be absorbed by the counterterms. Indeed, for the loop integral, $\tau$ is just a parameter and $a(\tau)\Lambda$ has an equivalent effect to a purely comoving cutoff.

To conclude the analysis of the quartic contribution we stress that it is indistinguishable from that of the mass counterterm, which means that it can be completely absorbed by the latter.

Let us now discuss the cubic contribution to the tensor spectrum; consider eq.\ \eqref{eq:DimReg c Ps} with $\delta = 0$, once more taking care of integrating first in momentum. 
Since the integration times $\tau'$ and $\tau''$ in that expression have a complex part derived from the $i\omega$ prescription, this introduces a damping in the UV limit of the momentum integrals which makes the resulting integral to be directly finite, see also \cite{Ballesteros:2024zdp}. Therefore, there is no need to introduce any kind of regulator in the momentum integrals.
This does not mean that the cubic contribution to the power spectrum is automatically finite: UV divergences are found in the time integrals when $\tau'\to\tau''$,\footnote{ Small spatial and temporal distances are equivalent to large momenta and frequencies in Fourier space. Thus, imposing a cutoff in momenta (and frequencies) is equivalent to imposing a cutoff on the smallest allowed distances in real space(time).} which physically requires regulating the integrals in eq.\ \eqref{eq:DimReg c Ps} as follows: 
\begin{align} \label{TI1int}
	\int_{-\infty_-}^\tau \dd \tau'  \int_{-\infty_+}^\tau \dd \tau'' \to \int_{-\infty_-}^{\tau - 1/(a(\tau)\Lambda)} \dd \tau'  \int_{-\infty_+}^\tau \dd \tau''  \,,
	\end{align}
	and
\begin{align} \label{TI2int}
	\int_{-\infty_-}^\tau \dd \tau'  \int_{-\infty_-}^{\tau'} \dd \tau'' \to \int_{-\infty_-}^\tau \dd \tau'  \int_{-\infty_-}^{\tau' - 1/(a(\tau')\Lambda)} \dd \tau''\,,
\end{align}
where we have again employed the most natural prescription to choose the times at which we evaluate $a(\tau)$.
In the first kind of double integral we encounter, eq.\ \eqref{TI1int}, we can do the integral in $\tau''$ obtaining a finite result, while a UV divergence is found in the upper limit of the integral in $\tau'$ if unregulated; we thus change the integration limit $\tau \to \tau - 1/ (a(\tau) \Lambda)$. In the second kind of double integral that we have to face, eq.\ \eqref{TI2int}, we find the UV divergence in the upper limit of the integral in $\tau''$; and  now we change the integration limit $\tau' \to \tau' - 1/ (a(\tau') \Lambda)$, being the integral in $\tau'$ finite. 
Although a priori the cubic contribution to the power spectrum is a real quantity, which is guaranteed by the exchange symmetry of this contribution in the $(\tau',\tau'')$ plane (see eq.\ \eqref{eq:2-Point 2}), with this regularization method we break this symmetry and we need to impose that this (cubic) contribution is real by hand.
As with the quartic contribution, other ways of regularizing the UV with a cutoff --e.g.\ changing the times at which we evaluate $a(\tau)$ in the regulator-- give rise to divergences that cannot be absorbed by the counterterms.

The final result for the cubic contribution to the tensor spectrum, $\mathcal{P}_h^{\rm 1l,c}$, once we absorb all divergences (except the logarithmic ones) and also the finite parts that can be absorbed with the counterterms, is:
\begin{equation}
	\mathcal{P}_h^{\rm 1l,c}(\tau,k) = \frac{3}{80} \frac{H^4}{\pi^4 M_P^4} \left(1+(k\tau)^2\right)\log\frac{k}{a(\tau) \Lambda}\,.
\end{equation}
Whereas the cubic contribution in dimensional regularization, eq.\ \eqref{conts3}, featured a factor of 2 difference between the coefficients of the $\log k$ and $1/\delta$ terms, now there is no such discrepancy between the logs of $k$ and $\Lambda$. 

Putting together all contributions (tree-level, quartic and cubic loops, and counterterms), absorbing the divergences and finite parts (when possible) with the counterterms, the result for the total tensor power spectrum is
\begin{align} \label{eq:tensorPS cutoff final}
	\nonumber
	\mathcal{P}_h(\tau,k) & = \frac{H^2}{M_P^2 \pi^2} \left(1+(k\tau)^2 \right) +\frac{3}{80} \left(  \frac{H^2}{M_P^2 \pi^2} \right) ^2  \bigg[ \alpha + \beta (k\tau)^2+\gamma(k\tau)^4 + \left(1+(k\tau)^2 \right) \log\frac{k}{\mu a(\tau)}\\
	& \quad\quad\quad\quad\quad\quad\quad\quad\quad\quad\quad\quad   + \kappa \Re{e^{2ik\tau}(i+k\tau)^2 \left(-i\pi+{\rm Ei}(-2ik\tau) \right)} \bigg] + \order{\frac{H^2}{M_P^2 \pi^2}}^3\,.
\end{align}
This result is the same as the one obtained by dimensional regularization (see eq.\ \eqref{eq:tensorPS}) once we impose that $\kappa = 1$. 
We stress the importance of having $\kappa = 1$ to avoid a late-time divergence. That such a divergence is avoided is reassuring.\footnote{As we discussed, such a divergence could be interpreted as the appearance of the generation of a late-time graviton mass.} 
The absence of late-time divergences arises naturally in dimensional regularization, 
whereas with a cutoff it is necessary to impose it a posteriori.
In practice, one can simply impose that the power spectrum cannot exhibit a behavior associated with having a late-time divergence to fix the constant $\kappa$ without the need for a comparison with dimensional regularization. An interesting analogy exists with QFT in flat spacetime: it is well known that in quantum electrodynamics (see Appendix \ref{sec:QED}) the use of a cutoff breaks gauge invariance, but loop-level  observables do not, thanks to the use of adequate counterterms. 

\section{Discussion} \label{sec:Conclusions}

We have presented a method for solving loop integrals in dimensional regularization by splitting them into a UV (divergent) piece that gets appropriately regularized and an IR (finite) piece that can be computed directly. The method is suited for cosmological applications, where loop integrals are notably more complex than in flat spacetime.

Working in $3+\delta$ spatial dimensions, UV logarithmic divergences are identified with single poles around $\delta = 0$ --see eq.\ \eqref{eq:Finite parts dim reg}--, while higher-order divergences are promoted to poles in $\delta\neq 0$ that disappear in the limit $\delta\to 0$, as usual. 
Finite contributions to loop-level in-in correlators arise in two ways. First, we obtain a finite contribution equivalent to the one we would obtain working in $3$ spatial dimensions by subtracting one by one all the UV divergences. And second, we also obtain a finite contribution that comes from the {interactions and the} mode functions evolving in $3+\delta$ spatial dimensions. 
Due to the $\order{\delta}$ suppression that this second contribution involves, only the coefficient accompanying the logarithmic divergence, ${\dd c_{-1}}/{\dd \delta}\eval_{\delta = 0}$,  enters in the finite result, as highlighted in \cite{Senatore:2009cf}.

We have illustrated the method with the calculation of the one-loop tensor power spectrum induced by scalar fluctuations during slow-roll inflation, $\mathcal{P}_h$, neglecting $\epsilon$ corrections and therefore effectively performing a computation in a de Sitter background. 

To complete the renormalization process, it is essential to include the effect of the counterterms. Using diffeomorphism invariance, we constructed the covariant action of counterterms for the tensor power spectrum, eq.\ \eqref{eq:Cts action general}, and expanding in fluctuations we find the operators that renormalize it, see eq.\ \eqref{eq:Cts general cov}. Although those counterterms are the ones that respect the original symmetries, regularization methods that break them (e.g.\ a cutoff) require new counterterms; specifically a mass term for tensor fluctuations.

In the context of dimensional regularization, we stress that obtaining the correct result requires working consistently in $3+\delta$ spatial dimensions, modifying not only the dynamics of the free fields but each and every contribution to the correlator of interest.

Given that the coefficients of the counterterms are in charge of the renormalization of the divergences, they decompose as $c = {c^{\rm div}}/{\delta} + c^{\rm fin}$, where $c^{\rm div}$ and $c^{\rm fin}$ are constants. In practice, the term of the power spectrum $\propto c^{\rm div}$ coming from the counterterms will give a divergent $\sim 1/\delta$ piece and a finite contribution, being the latter sensitive to the physics in $3+\delta$ spatial dimensions. This finite contribution is intimately related to the structure of the UV logarithmic divergences, and is as important as the logarithmic running that the one-loop contribution induces on the power spectrum. 

The need to include the correct counterterms becomes clear taking into consideration that an incorrect choice may also allow to absorb the one-loop divergences, but their finite contribution to the correlator being renormalized may be radically different, leading to erroneous predictions for observable quantities. We have discussed this point with a concrete example (the computation of $\mathcal{P}_h$); this observation, however, extends in general to any correlator at loop level. 

The tensor power spectrum $\mathcal{P}_h$ that we have obtained, eq.\ \eqref{eq:tensorPS}, is a function of comoving momentum and conformal time valid while (slow-roll) inflation lasts and is invariant under the dilatations. The tensor power spectrum at one loop does {\it not} feature
a late-time divergence $\log a(\tau)$. Its absence becomes manifest
once the cubic and quartic contributions of the one-loop, as well as the contribution of the counterterms, are included. To realize the cancellation of the sum of these late-time divergences, the inclusion of the full dynamics of the fields in $3+\delta$ spatial dimensions is essential. This highlights the importance of having a tool that allows to obtain the finite parts of the loop integrals in dimensional regularization.

We have also calculated $\mathcal{P}_h$ regularizing with a physical (as opposed to comoving) cutoff, see also \cite{Senatore:2009cf}. As mentioned above, when using a cutoff, new counterterms that violate diffeomorphism invariance (such as a mass term) are required. These counterterms make the result dependent on more constants than those required for a regularization method such as dimensional regularization. However, imposing that the final power spectrum is free of late-time divergences, the result obtained coincides exactly with that of dimensional regularization for all times $\tau$. As it is often the case, we find that it is convenient to use a regulator that respects the symmetries so that the final observable is the correct one without having to impose additional conditions.

We expect that the method we have discussed to compute loop integrals in dimensional regularization will be useful to deal with other QFT problems in cosmology and may also be applied in other contexts.

\vspace{0.3cm}

\mysection{Acknowledgments}

{\small
We thank the organizers of the CERN-TH Institute ``Looping in the Primordial Universe'', where this work was presented by FR on October 29, 2024. 
We thank G.\ Franciolini and J.\ Garriga for discussions.\\ Work funded by the following grants: PID2021-124704NB-I00 funded by MCIN/AEI/10.13039/501100011033 \sloppy and by ERDF A way of making Europe, CNS2022-135613 MICIU/AEI/10.13039/501100011033 and by the European Union NextGenerationEU/PRTR, and Centro de Excelencia Severo Ochoa CEX2020-001007-S funded by MCIN/AEI/10.13039/501100011033.
JGE is supported by a PhD contract {\it contrato predoctoral para formaci\'on de doctores} (PRE2021-100714) associated to the aforementioned Severo Ochoa grant, CEX2020-001007-S-21-3. 
}

\appendix

\section{Bubbles and disconnected parts} \label{ap: Bubbles and disconnected parts}

In this section we show that bubble diagrams can be left out when computing correlators using the in-in formalism. We also show that the disconnected parts of the $n$-point correlators can be obtained by connected correlators of lower orders. Starting from the in-in formula in the operator formalism,
\begin{equation}
	\expval{\mathcal{O}(t)} = \frac{\bra{0} F^{-1}(t,-\infty_+) \mathcal{O}_I(t)F(t,-\infty_-) \ket{0}}{\bra{0} F^{-1}(t,-\infty_+) F(t,-\infty_-) \ket{0}} 
\end{equation}
and expanding the time evolution operator in the interaction picture,
\begin{equation}
	F(t,-\infty_-) = \sum_{n = 0}^\infty \frac{(-i)^n}{n!} \int_{-\infty_-}^t \dd t_1\mydots\dd t_n\, T\left\lbrace H_I(t_1)\mydots H_I(t_n) \right\rbrace 
\end{equation}
we obtain that:
\begin{align}\label{eq:In-In correlation expanded}
	\nonumber
	\bra{0} F^{-1}(t,-\infty_+) \mathcal{O}_I(t)F(t,-\infty_-) \ket{0} & = \sum_{n,m = 0}^\infty \frac{(-i)^n i^m}{n!m!} \int_{-\infty_-}^t \dd t_1\mydots\dd t_n \int_{-\infty_+}^t \dd \tilde{t}_1\mydots\dd \tilde{t}_m\\
	& \times \bra{0} \overline{T}\left\lbrace H_I(\tilde{t}_1)\mydots H_I(\tilde{t}_m) \right\rbrace \mathcal{O}_I(t) T\left\lbrace H_I(t_1)\mydots H_I(t_n) \right\rbrace \ket{0}\,.
\end{align}
Given the correlation
\begin{align}\label{eq:In-In correlation expanded aux}
	\int_{-\infty_-}^t \dd t_1\mydots\dd t_n \int_{-\infty_+}^t \dd \tilde{t}_1\mydots\dd \tilde{t}_m	\bra{0} \overline{T}\left\lbrace H_I(\tilde{t}_1)\mydots H_I(\tilde{t}_m) \right\rbrace \mathcal{O}_I(t) T\left\lbrace H_I(t_1)\mydots H_I(t_n) \right\rbrace \ket{0}\,,
\end{align}
we will distinguish between terms that are bubbles and terms that are not. A bubble contribution is such that there is a subset of the interaction Hamiltonians in the interaction picture, $H_I$, that are not connected by correlations with the operator $\mathcal{O}_I$. Since all the fields appearing in eq.\ \eqref{eq:In-In correlation expanded aux} are in the interaction picture, eq.\ \eqref{eq:In-In correlation expanded aux} is calculated following the Wick's theorem \cite{Wick:1950ee}: all possible correlations of the fields will be taken two by two, being the final result the sum of all possible combinations. Thus, given eq.\ \eqref{eq:In-In correlation expanded aux}, there will be a contribution where all $H_I$ are connected to $\mathcal{O}_I$, another where one $H_I$ is not connected to $\mathcal{O}_I$, and so on until no $H_I$ is connected to $\mathcal{O}_I$. We will return to this point later, but first we will make an important clarification: if a subset with cardinal $n'$ of $H_I$ within $T\left\lbrace H_I(t_1)\mydots H_I(t_n) \right\rbrace$ is not connected to the rest of the $H_I$ their order does not matter and we can separate the time-ordering operators:
\begin{equation}
	T\left\lbrace H_I(t_1)\mydots H_I(t_n) \right\rbrace = T\left\lbrace H_I(t_1)\mydots H_I(t_{n'}) \right\rbrace \times T\left\lbrace H_I(t_{n'+1})\mydots H_I(t_n) \right\rbrace\,.
\end{equation}
An analogous equation applies to ${\overline T}$. Since within time-ordering the position of $H_I$ is irrelevant, we have $\binom{n}{n'}= n!/(n'!(n-n')!)$ independent ways of taking groups of $n'$ Hamiltonians within the original set of $n$.
Using this result,
\begin{align}
	\nonumber
	&\int_{-\infty_-}^t \dd t_1\mydots\dd t_n \int_{-\infty_+}^t \dd \tilde{t}_1\mydots\dd \tilde{t}_m	\bra{0} \overline{T}\left\lbrace H_I(\tilde{t}_1)\mydots H_I(\tilde{t}_m) \right\rbrace \mathcal{O}_I(t) T\left\lbrace H_I(t_1)\mydots H_I(t_n) \right\rbrace \ket{0} =\\
	\nonumber
	& = \int_{-\infty_-}^t \dd t_1\mydots\dd t_n \int_{-\infty_+}^t \dd \tilde{t}_1\mydots\dd \tilde{t}_m	\sum_{n'=0}^n  \sum_{m'=0}^m \binom{n}{n'} \binom{m}{m'}  \bra{0} \overline{T}\left\lbrace H_I(\tilde{t}_1)\mydots H_I(\tilde{t}_{m'}) \right\rbrace T\left\lbrace H_I(t_1)\mydots H_I(t_{n'}) \right\rbrace \ket{0}\\
	& \quad\times \bra{0} \overline{T}\left\lbrace H_I(\tilde{t}_{m'+1})\mydots H_I(\tilde{t}_m) \right\rbrace \mathcal{O}_I(t) T\left\lbrace H_I(t_{n'+1})\mydots H_I(t_n) \right\rbrace \ket{0}\eval_\textrm{no bubbles}\,.
\end{align}
We have separated the packets $\overline{T}\left\lbrace H_I(\tilde{t}_1)\mydots H_I(\tilde{t}_m) \right\rbrace$ and $T\left\lbrace H_I(t_1)\mydots H_I(t_n) \right\rbrace$ into as many as possible two smaller subpackets. Without loss of generality, we consider the first packet as purely disconnected from $\mathcal{O}_I$ and the second as purely connected to $\mathcal{O}_I$ (i.e.\ it will have no bubbles). We have used the symmetry $t_i\leftrightarrow t_j$ and $\tilde{t}_i \leftrightarrow \tilde{t}_j$ (we can always redefine the integration variables) to
note that each of the ways of making the $m'$ subpackets within $m$ (and $n'$ within $n$) will contribute equally to the correlation, generating the $\binom{n}{n'} \binom{m}{m'}$ binomial factors. We stress that the above equation contains all possible contributions to the correlation function, with the advantage that we have explicitly separated the bubble terms from the non-bubble ones. Finishing the proof is simple: we just plug this result into the expression \eqref{eq:In-In correlation expanded},
\begin{align}
	\nonumber
	&\bra{0} F^{-1}(t,-\infty_+) \mathcal{O}_I(t)F(t,-\infty_-) \ket{0} = \sum_{n,m = 0}^\infty \sum_{n'=0}^n \sum_{m'=0}^m \frac{(-i)^{n'} i^{m'}}{n'!m'!} \frac{(-i)^{n-n'} i^{m-m'}}{(n-n')!(m-m')!} \\
	\nonumber &\times\int_{-\infty_-}^t \dd t'_1\mydots\dd t'_{n-n'} \int_{-\infty_+}^t \dd \tilde{t}'_1\mydots\dd \tilde{t}'_{m-m'} \bra{0} \overline{T}\left\lbrace H_I(\tilde{t}'_1)\mydots H_I(\tilde{t}'_{m-m'}) \right\rbrace \mathcal{O}_I(t) T\left\lbrace H_I(t'_1)\mydots H_I(t'_{n-n'}) \right\rbrace \ket{0}\eval_{{\rm n.b.}}\\
	&\times\int_{-\infty_-}^t \dd t_1 \mydots \dd t_{n'} \int_{-\infty_+}^t \dd \tilde{t}_1\mydots\dd \tilde{t}_{m'} \bra{0} \overline{T}\left\lbrace H_I(\tilde{t}_1)\mydots H_I(\tilde{t}_{m'}) \right\rbrace T\left\lbrace H_I(t_1)\mydots H_I(t_{n'}) \right\rbrace \ket{0}\,,
\end{align}
and note that the sums are factorized so that for each value of $n'$ and $m'$ fixed, the whole tower $n-n' = 0,1,\dots$ and $m-m' = 0,1,\dots$ is generated, so that
\begin{align}
	\nonumber
	\bra{0} F^{-1}(t,-\infty_+) \mathcal{O}_I(t)F(t,-\infty_-) \ket{0} & = \bra{0} F^{-1}(t,-\infty_+) F(t,-\infty_-) \ket{0}\\
	& \times \bra{0} F^{-1}(t,-\infty_+) \mathcal{O}_I(t)F(t,-\infty_-) \ket{0}\eval_\textrm{no bubbles}\,.
\end{align}
We thus conclude that the in-in formula in the operator formalism can be written as
\begin{equation}
	\expval{\mathcal{O}(t)} =\bra{0} F^{-1}(t,-\infty_+) \mathcal{O}_I(t)F(t,-\infty_-) \ket{0}\eval_\textrm{no bubbles}\,.
\end{equation}

The proof that disconnected $n$-point correlation functions are generated by connected correlations of lower order is very similar. We start by assuming the {following form for the} operator $\mathcal{O}(t) = \Phi_1(t,\vb{k}_1) \mydots \Phi_n(t,\vb{k}_n)$, being $\Phi_i$ completely {generic} fields. As usual, we refer to each of the fields appearing in $\mathcal{O}(t)$ as external legs. We are interested in calculating
\begin{align}
	\nonumber
	&\expval{\Phi_1(t,\vb{k}_1) \mydots \Phi_n(t,\vb{k}_n)} = \sum_{n,m = 0}^\infty \frac{(-i)^n i^m}{n!m!} \int_{-\infty_-}^t \dd t_1\mydots\dd t_n \int_{-\infty_+}^t \dd \tilde{t}_1\mydots\dd \tilde{t}_m\\
	&\hspace{2.4cm} \times \bra{0} \overline{T}\left\lbrace H_I(\tilde{t}_1)\mydots H_I(\tilde{t}_m) \right\rbrace \Phi_1^I(t,\vb{k}_1) \mydots \Phi_n^I(t,\vb{k}_n) T\left\lbrace H_I(t_1)\mydots H_I(t_n) \right\rbrace \ket{0} \eval_\textrm{no bubbles}\,.
\end{align}
For convenience, we will define
\begin{align}
	\nonumber
	&{(n,m:\Phi_1(t,\vb{k}_1),\Phi_2(t,\vb{k}_2), \dots)} \equiv \int_{-\infty_-}^t \dd t_1\mydots\dd t_n \int_{-\infty_+}^t \dd \tilde{t}_1\mydots\dd \tilde{t}_m\\
	&\hspace{4.1cm} \times\bra{0} \overline{T}\left\lbrace H_I(\tilde{t}_1)\mydots H_I(\tilde{t}_m) \right\rbrace \Phi_1^I(t,\vb{k}_1) \Phi_2^I(t,\vb{k}_2) \mydots T\left\lbrace H_I(t_1)\mydots H_I(t_n) \right\rbrace \ket{0}_c \,.
\end{align}
The subindex $_c$ ({\it connected}) imposes that all elements within the correlation are linked in a connected way. Of course this guarantees the non-existence of bubbles, but even more it imposes the connection between all external legs. All possible contractions in this correlation will be taken into account within these constraints. Thus, the one-point correlation --taking into account that bubbles do not
contribute as we have shown above-- can only be
\begin{equation}
	\expval{\Phi_1(t,\vb{k}_1)} = \sum_{n,m = 0}^\infty \frac{(-i)^n i^m}{n!m!} (n,m:\Phi_1(t,\vb{k}_1))\,.
\end{equation}
For two-point correlation we have two possibilities: the two external legs may (or may not) be connected. In the case where they are not connected, we must repeat the procedure used in the {previous proof for the irrelevance of bubbles}: we must separate $T$ and $\overline{T}$ into two subpackets, each of these dedicated to each of the external legs. We thus have that:
\begin{align}
	\nonumber
	\expval{\Phi_1(t,\vb{k}_1) \Phi_2(t,\vb{k}_2)} & = \sum_{n,m = 0}^\infty \frac{(-i)^n i^m}{n!m!} \bigg\lbrace (n,m:\Phi_1(t,\vb{k}_1),\Phi_2(t,\vb{k}_2)) \\
	& + \sum_{n'=0}^n \sum_{m'=0}^m \binom{n}{n'} \binom{m}{m'} (n',m':\Phi_1(t,\vb{k}_1)) (n-n',m-m':\Phi_2(t,\vb{k}_2)) \bigg\rbrace \,.
\end{align}
After working out the second line, as in the case of the bubbles, the contributions are factorized in such a way that
\begin{equation}
	\expval{\Phi_1(t,\vb{k}_1) \Phi_2(t,\vb{k}_2)} = \expval{\Phi_1(t,\vb{k}_1) \Phi_2(t,\vb{k}_2)}_c + \expval{\Phi_1(t,\vb{k}_1)}_c \expval{\Phi_2(t,\vb{k}_2)}_c \,,
\end{equation}
i.e.\ the disconnected part of the two-point correlation is obtained from the product of the connected parts of the one-point correlations.
Following this procedure, it is straightforward to note that, for any number of fields: 
\begin{align}
	\nonumber
	\expval{\Phi_1(t,\vb{k}_1) \Phi_2(t,\vb{k}_2)\mydots} = \expval{\Phi_1(t,\vb{k}_1) \Phi_2(t,\vb{k}_2)\mydots}_c + \big(  \expval{\Phi_1(t,\vb{k}_1)}_c \times\expval{\Phi_2(t,\vb{k}_2)\dots}_c + {\rm perms.} \big) \\
	+ \dots +  \expval{\Phi_1(t,\vb{k}_1)}_c \times \expval{\Phi_2(t,\vb{k}_2)}_c \times \mydots \,.
\end{align}
The $n$-point correlation is generated by all possible connected m-point correlations, with $m\leq n$.

\section{QED analogy} \label{sec:QED}

The analysis we have done for the tensor power spectrum $\mathcal{P}_h$ induced by scalar modes shows how, although there is no late-time divergence using dimensional regularization, when a cutoff is used as a regulator this condition has to be imposed a posteriori to obtain the correct spectrum. In this section, we will perform an analogous analysis in {quantum electrodynamics} 
to better understand the nature of this result. In particular, we are going to analyze the one-loop photon propagator, known as vacuum polarization,
\begin{equation}
	\begin{tikzpicture}[baseline={-2}]
		\draw[decorate, decoration=snake] (5,0) -- (6.5,0);
		\draw[decorate, decoration=snake] (7.5,0) -- (9,0);
		\draw[postaction={decorate}, decoration={markings, mark=at position 0.5 with {\arrow{<}}}] 
		    (7.5,0) arc[start angle=0, end angle=180, radius=0.5];
		\draw[postaction={decorate}, decoration={markings, mark=at position 0.5 with {\arrow{<}}}] 
		    (6.5,0) arc[start angle=180, end angle=360, radius=0.5];
		
		\fill[black] (5,0) circle (1.5pt);
		\fill[black] (6.5,0) circle (1.5pt);
		\fill[black] (7.5,0) circle (1.5pt);
		\fill[black] (9,0) circle (1.5pt);
		
		\node[black] at (5,0.3) {$\mu$};
		\node[black] at (9,0.3) {$\nu$};
		
		\node[black] at (5.75,0.3) {$q$};
		
		\node[black] at (5.75,-0.3) {$\gamma$};
		\node[black] at (7,-0.8) {$e$};
		
		\node[black] at (10.2,0) {$\equiv i \Pi_2^{\mu\nu}(q)\,,$};	
	\end{tikzpicture}
\end{equation}
which, calculated using the usual methods, acquires the form
\begin{align}
	\label{eq:QED initialintegral}
	i\Pi_2^{\mu \nu}(q) = -4ie^2\int_0^1 dx \int \frac{d^4 \ell_E}{(2\pi)^4}\frac{\frac{1}{2}g^{\mu\nu}\ell_E^2-2x(1-x) q^\mu q^\nu+g^{\mu\nu}[m^2+x(1-x)q^2]}{[\ell_E^2+m^2-x(1-x)q^2]^2}\,.
\end{align}
This one-loop contribution to the photon propagator is divergent in the UV, so it must be regularized and renormalized. We will {use} dimensional regularization and a cutoff.
Thanks to the fact that the theory has a gauge symmetry, the vacuum polarization is constrained by the Ward identities,
\begin{equation}\label{eq:QED WI}
	q_\mu\Pi_2^{\mu \nu}(q)=0\ , \quad \textrm{which ensures that}\quad \Pi_2^{\mu \nu}(q) = \Pi_2(q^2)(q^2g^{\mu\nu}-q^\mu q^\nu)\,,
\end{equation}
which is related to the massless nature of the photon.

\subsection*{Dimensional regularization}

In dimensional regularization we generalize the equation \eqref{eq:QED initialintegral} for $d=4-\delta$ spacetime dimensions:
\begin{align}
	i\Pi_2^{\mu \nu}(q) = -4ie^2\int_0^1 dx \int \mu^\delta \frac{d^d \ell_E}{(2\pi)^4}\frac{\left(1-\frac{2}{d}\right)g^{\mu\nu}\ell_E^2-2x(1-x) q^\mu q^\nu+g^{\mu\nu}[m^2+x(1-x)q^2]}{[\ell_E^2+m^2-x(1-x)q^2]^2}.
\end{align}
Since dimensional regularization preserves gauge invariance, the structure of the vacuum polarization imposed by the Ward identities \eqref{eq:QED WI} is guaranteed, obtaining:
\begin{align} \label{eq:divergentpropagatordimreg}
	\nonumber
	i\Pi_2(q^2) = -\frac{ie^2}{2\pi^2} & \Bigg\{ \frac{1}{6} \left[ \frac{2}{\delta} - \gamma_E + \log(4\pi) \right] + \frac{5}{18} + \frac{2m^2}{3q^2}\\
	& - \frac{\sqrt{4m^2-q^2}}{3q^3} (2m^2+q^2) \arctan \left( \frac{q} {\sqrt{4m^2-q^2}} \right) + \frac{1}{6}\log\left(\frac{\mu^2}{m^2}\right)\Bigg\}\,,
\end{align}
where we are restricting to the case $q^2 < (2m)^2$ to avoid the branch cut. To renormalize the divergence of $\Pi_2(q^2)$, expressed by the simple pole in $\delta = 0$, we need to resort to the counterterm Lagrangian generated by the wave function of the photon, 
\begin{align}
	\label{eq:counterterm}
	\mathcal{L}_{\rm cts}=-\frac{1}{4}(Z_3-1)F_{\mu \nu}F^{\mu \nu} = - \frac{1}{2}(Z_3-1) \left(\partial_\mu A_\nu \partial^\mu A^\nu-\partial_\mu A_\nu \partial^\nu A^\mu \right) \,,
\end{align}
which induces the Feynman rule 
\begin{equation}
	\begin{tikzpicture}[baseline={-2}]
		\draw[decorate, decoration=snake] (0,0) -- (2,0);
		
		\fill[black] (0,0) circle (1.5pt);
		\fill[black] (2,0) circle (1.5pt);
		
		\node[black] at (0,0.3) {$\mu$};
		\node[black] at (2,0.3) {$\nu$};
		
		\draw[fill=white,cross] (1,0) circle (0.2);
		\node[black] at (0.5,0.3) {$q$};
		
		\node[black] at (4.7,0) {$= -i(Z_3-1)(g^{\mu\nu}q^2-q^\mu q^\nu)\,.$};	
	\end{tikzpicture}
\end{equation}
We can use $Z_3$ to absorb not only the divergent part of $\Pi_2(q^2)$, but also the finite numerical coefficients, obtaining that
\begin{align}
	\nonumber
	i\Pi_2(q^2) = -\frac{ie^2}{2\pi^2} & \Bigg[\alpha + \frac{2m^2}{3q^2} - \frac{\sqrt{4m^2-q^2}}{3q^3} (2m^2+q^2) \arctan \left( \frac{q} {\sqrt{4m^2-q^2}} \right) \\
	&+\frac{1}{6}\log\left(\frac{\mu^2}{m^2}\right)\Bigg] \equiv i\Pi_2(q^2)^{\rm dim.reg.}_{\rm ren}\,,
\end{align}
where $\alpha$ is the finite contribution of the counterterms to be determined by the experiment. 

\subsection*{Cutoff regularization}
Regularizing the integral \eqref{eq:QED initialintegral} with a cutoff $\Lambda^2 > \ell_E^2 $ in $d = 4$ dimensions, we get
\begin{align} \label{eq:divergentpropagatorcutoff}
	\nonumber
	i\Pi_2^{\mu\nu}(q) & = \left[ -\frac{ie^2}{2\pi^2}\left(\frac{1}{6}\log\left(\frac{\Lambda^2}{\mu^2}\right)+\frac{1}{9}\right) + i\Pi_2(q^{2})^{\rm dim.reg.}_{\rm ren}\eval_{\alpha = 0} \right] (q^2g^{\mu\nu}-q^\mu q^\nu)\\
	&-\frac{ie^2}{2\pi^2}\left(\frac{\Lambda^2-m^2}{4}+\frac{q^2}{24}\right)g^{\mu\nu}\,.
\end{align}
The tensor structure of the second line does not respect the Ward identity described in \eqref{eq:QED WI} because the cutoff breaks gauge invariance. 
This implies that when constructing the counterterm Lagrangian, interactions that were previously forbidden by gauge symmetry now contribute to the photon propagator. In particular, we can introduce a mass term for the photon, as well as more freedom in the kinetic term:
\begin{align}
	\label{eq:QED counterterm cutoff}
	\mathcal{L}_{\rm cts}= \frac{c}{2} \partial_\mu A_\nu \partial^\mu A^\nu + \frac{c'}{2} \partial_\mu A_\nu \partial^\nu A^\mu + \frac{1}{2}M^2 A_\mu A^\mu  \,,
\end{align}
which induces the diagram
\begin{align}
	\begin{tikzpicture}[baseline={-2}]
		\draw[decorate, decoration=snake] (0,0) -- (2,0);		
		\fill[black] (0,0) circle (1.5pt);
		\fill[black] (2,0) circle (1.5pt);	
		\node[black] at (0,0.3) {$\mu$};
		\node[black] at (2,0.3) {$\nu$};	
		\draw[fill=white,cross] (1,0) circle (0.2);
		\node[black] at (0.5,0.3) {$q$};
		\node[black] at (5,0) {$= i \left(c\, g^{\mu\nu}q^2 + c'\, q^\mu q^\nu + M^2g^{\mu\nu} \right) \,.$};	
	\end{tikzpicture}
\end{align}
We can use these counterterms to absorb the divergences (and finite parts) of the vacuum polarization, obtaining
\begin{align} \label{eq:QED cutoff final}
	i\Pi_2^{\mu\nu}(q)^{\rm cutoff}_{\rm ren} = i\Pi^{\mu\nu}_2(q)_{\rm ren}^{\rm dim.reg.} -\frac{ie^2}{2\pi^2}\left(\beta +\gamma\, q^2 \right)g^{\mu\nu}\,,
\end{align}
where, just like $\alpha$, now $\beta$ and $\gamma$ are finite contributions of the counterterms a priori to be determined by experiment. 
We note that, in order to obtain the result calculated with dimensional regularization, we must impose that $\beta = 0$ and $\gamma = 0$. However, we can obtain the correct vacuum polarization using a cutoff by imposing at the end of the calculation that gauge invariance must be respected, i.e.\ by imposing that $q_\mu \Pi^{\mu\nu} = 0$, without the need to go through dimensional regularization. We stress that both results, the one obtained by means of dimensional regularization and the one with cutoff, are compatible thanks to the fact that in this second case new counterterms arise which are able to make disappear the gauge artifacts arising in the last term of the  equation \eqref{eq:QED cutoff final}. The analogy with the computation of the tensor power spectrum is clear: using dimensional regularization the late-time divergence canceled in a non-trivial way, while using a cutoff it did not. However, using a cutoff we were able to introduce a mass counterterm which, although a priori introduced an additional constant in the final power spectrum (the constant $\kappa$ of eq.\ \eqref{eq:tensorPS cutoff}), it was determined under the condition that the late-time divergence should vanish. 

\section{In-out comparison} \label{ap:in-out}

In this appendix we check the consistency between renormalization in the in-in and the in-out formalisms, for the action we have used throughout this work --see Section \ref{sec:Action}--, focusing on the one-loop power spectrum of tensor modes (in-in) and their propagator (in-out). The divergent parts of counterterms are set under the condition that observables constructed in any formalism of the two are finite. This should guarantee that observables constructed in the other formalism are automatically finite as well, since the action describing both types of observables is the same. Although in flat spacetime this last statement could be easily verified,
on a curved background there are essential difficulties with the definition of the in-out observables, see e.g\ \cite{Bousso:2004tv}. For instance, trying to construct an S-matrix in de Sitter, one faces the difficulty that the asymptotic states cannot be defined in the past and future conformal space-like boundaries.\footnote{See \cite{Melville:2023kgd} for a recent proposal to define an S-matrix in de Sitter and \cite{Mirbabayi:2022gnl} for a calculation of an in-out correlator with the fields pushed to the future and past horizons of a comoving observer.}

Considering inflation, a difficulty for correctly defining observable quantities in the in-out formalism resides in the fact that, taking as {\it out} time the end of inflation, $\tau_0 \to 0^-$, the asymptotic vacuum is far from being free, violating the foundations of the in-out formalism (see e.g.\ \cite{Donath:2024utn}).\footnote{We cannot go from the vacuum in the Heisenberg picture $\ket{\Omega}$ to the vacuum in the interaction picture $\ket{0}$ by simply adding the $i\omega$ prescription. At the beginning of inflation, all the modes of interest are well inside the horizon and their evolution is that of a free field in flat spacetime, except for canonical normalization factors --e.g. the term $1/a(\tau)$ for a scalar field--, which is what allows us to transform $\ket{\Omega} \to \ket{0}$. At the end of inflation, this is no longer true.}
However, we are going to argue that one can take simplifying limits that allow to recover from the in-out perspective the same renormalization conditions that we found imposing finiteness of the (in-in) power spectrum.

Let us take the following ansatz for the Feynman propagator of the tensor field in the in-out formalism:
\begin{align}
    \Pi_{ijkl}(x,y) = \bra{0} T\left\{h_{ij}(x)h_{kl}(y)e^{-i\int_{-\infty_-}^{\tau_{0,+}} H_{I}(\tau)\dd \tau}\right\}\ket{0}\bigg|_{\textrm{no bubbles}}\,,
\end{align}
where all the fields are meant to be in the interaction picture.\footnote{Although in flat spacetime the upper limit of the time integral --where the theory becomes asymptotically free-- is $t\to \infty (1-i\omega)$, during inflation, since $\tau_0<0$, to properly project onto the interaction vacuum the upper limit must be $\tau_0(1+i\omega)$, with positive $\omega$.} Expanding at second order in the interaction Hamiltonian, we find:
\begin{align}
    & \Pi^{(0)}_{ijkl}(x,y) = \bra{0} T\left\{h_{ij}(x)h_{kl}(y)\right\} \ket{0} \,,\\
    & \Pi^{(1)}_{ijkl}(x,y) =-i\int_{-\infty_-}^{\tau_{0,+}} \dd \tau \bra{0} T\left\{h_{ij}(x)h_{kl}(y)H_I(\tau)\right\}\ket{0}\,,\\
    & \Pi^{(2)}_{ijkl}(x,y) =-\frac{1}{2}\int_{-\infty_-}^{\tau_{0,+}} \dd \tau \int_{-\infty_-}^{\tau_{0,+}} \dd \tau' \bra{0} T\left\{h_{ij}(x)h_{kl}(y)H_I(\tau)H_I(\tau')\right\}\ket{0}\,.
\end{align}
Using the action described by eq.\ \eqref{eq:Action complete dim reg}, we see that the counterterms and the quartic interaction contribute to $\Pi^{(1)}_{ijkl}$, while the cubic interaction contributes to $\Pi^{(2)}_{ijkl}$. Working with the Fourier decomposition,
\begin{align}
    \Pi_{ijkl}(x,y)\equiv \int \frac{\dd^3 \vb{q}}{(2\pi)^3}e^{i\vb{q}\cdot(\vb{x}-\vb{y})}\sum_{\gamma=+,\times}e^\gamma_{ij}(\vb{q}) e^\gamma_{kl}(\vb{q}) \Pi_q(x^0,y^0)\,,
\end{align}
where $x^0$ and $y^0$, respectively, are the conformal times of $x$ and $y$ and $\vb{q}$ is the comoving momentum (with modulus $q$).
We find late-time divergences that appear as poles of various orders around $\tau_0 = 0$ (the end of inflation), an indication that the theory is not asymptotically free in this limit.
Restricting our analysis to the divergent part of the one-loop two-point correlation, i.e.\ extracting the coefficient of the pole $1/\delta$ in dimensional regularization, and assuming $x^0>y^0$ for simplicity, we find that its non-vanishing contribution comes from the cubic diagram and reads:
 \begin{align}
    \Pi_k^{\rm 1l,c}(x^0,y^0)\eval_{\rm divergent}= \frac{1}{\delta}\frac{3 H^4 }{40 \pi ^2  k^3 M_P^4} (1+i k x^0) (1-i k y^0) e^{-i k (x^0-y^0)}\,,
\end{align}
whereas the quartic diagram contribution to the pole vanishes:
\begin{align}
    \Pi_k^{\rm 1l,q}(x^0,y^0)\eval_{\rm divergent}= 0\,.
 \end{align}
As we have mentioned, modes of arbitrary comoving momentum do not behave as free fields in a flat spacetime  asymptotically. However, in the limit $k\to \infty$ the modes are well inside the Hubble radius and behave asymptotically as required --for fixed values of $\tau_0$-- according to the basics of in-out formalism, which suggests that it can be possible to extract information from $\Pi_k(x^0,y^0)$ in this limit. Therefore we impose the finiteness condition
\begin{align}
\label{eq:inoutfiniteness}
    \left(\Pi^{\rm 1l}_k+\Pi^{\rm cts}_k\right)\bigg|_{\rm divergent} = 0\,,
\end{align}
in the limit $k\to\infty$ through an expansion in powers of the comoving momentum. For the sake of brevity,\footnote{We stress that the computations summarized in this appendix to obtain the one-loop tensor propagator are not simpler than the ones described in the main body of this work to obtain the one-loop tensor spectrum.} we do not report here the counterterm contribution to the propagator, $\Pi^{\rm cts}_k$. At linear order in $k$, eq.~\eqref{eq:inoutfiniteness} takes the form
\begin{align}
    \frac{C_3^{\rm div} H^4 k x^0 y^0 e^{-i k (x^0-y^0)}}{M_P^4\,\delta} \left[(x^0)^2-(y^0)^2-e^{2ik(x^0-\tau_0)}\tau_0^2\right] =0\,, \quad \textrm{which implies} \quad C_3^{\rm div}=0\,.
\end{align}
At $\order{k^0}$, eq.\ \eqref{eq:inoutfiniteness} is
\begin{align}
    \frac{H^4 e^{-i k (x^0-y^0)}}{M_P^4\,\delta} x^0 y^0(x^0-y^0)(C_1^{\rm div}+C_2^{\rm div}) =0\,, \quad \textrm{implying} \quad C_1^{\rm div}=-C_2^{\rm div}\,.
\end{align}
Finally, at $\order{k^{-1}}$, we have
\begin{align}
    \frac{H^4e^{-i k (x^0-y^0)}}{k\, M_P^4\,\delta}x^0 y^0\left[320\pi^2C_2^{\rm div}(e^{2ik(x^0-\tau_{0,+})}-2)+3\right]=0\,.
\end{align}
Taking into account the prescription $\tau_{0,+}\to\tau_0(1+i\omega)$, and sending $k\to\infty$, we find $C_2^{\rm div}={3}/(640\pi^2)$.
In this way, we recover the same conditions on the divergent parts of the counterterms that we obtained with the in-in formalism (see eqs.\ \eqref{eq:Cts redef} and \eqref{eq:Cts sol}), as expected. 

\bibliography{draft}

\end{document}